\documentclass[3p,nopreprintline]{elsarticle}

\usepackage{graphicx}
\usepackage[pdfencoding=auto]{hyperref}
\hypersetup{%
 bookmarksnumbered=true,%
 colorlinks=true,%
 pdftitle={SuperKEKB Collider},%
 pdfauthor={Kazunori Akai, Kazuro Furukawa, Haruyo Koiso},%
 pdfsubject={NIM A doi:10.1016/j.nima.2018.08.017},%
 pdfkeywords={SuperKEKB\000\012B factory\000\012Asymmetric collider\000\012}}

\begin{document}

\begin{frontmatter}

\title{SuperKEKB Collider}

\author{Kazunori Akai, Kazuro Furukawa, Haruyo Koiso\\
on behalf of the SuperKEKB accelerator team}
\address{
High Energy Accelerator Research Organization (KEK), 1-1 Oho, Tsukuba, Ibaraki, 305-0801, Japan\\
SOKENDAI (The Graduate University for Advanced Studies), Hayama, Kanagawa 240-0193, Japan}

\hyphenation{SuperKEKB}

\begin{abstract}
SuperKEKB, a 7 GeV electron -- 4 GeV positron double-ring collider, is constructed by upgrading KEKB in order to seek new physics beyond the Standard Model.
The design luminosity of SuperKEKB is $8\times10^{35} \rm{cm^{-2}s^{-1}}$ --- 40 times higher than that achieved by KEKB. The greater part of the gain comes from significantly decreasing the beam sizes at the interaction point based on the nanobeam collision scheme; the design beam currents in both rings are double those achieved in KEKB.
Large-scale construction to upgrade both the collider rings and the injector was conducted, and beam commissioning without the Belle II detector and final-focus magnets was successfully carried out from February to June in 2016. 
Subsequently, renovation of the interaction region, including the installation of the final-focus magnets and Belle II, and construction in the final stage of a new positron damping ring 
were conducted. Having completed the interaction region, beam collision tuning 
is scheduled from March till July in 2018. This paper reviews the design, construction, and beam commissioning of SuperKEKB. 
\end{abstract}


\end{frontmatter}



\section{Introduction}

SuperKEKB is an asymmetric-energy electron--positron double-ring collider constructed by upgrading the KEKB B-Factory~\cite{KEKB}. 
KEKB, which operated from 1998 until June 2010, achieved the world's highest luminosity ($2.11\times10^{34} \rm{cm^{-2}s^{-1}}$); the total integrated luminosity accumulated by the Belle detector reached 1.04 $\rm{ab^{-1}}$~\cite{ptep-kekb}. 
Using these data, the Belle collaboration succeeded in proving the Kobayashi--Maskawa theory and obtained a variety of important experimental results in elementary particle physics. 
Based on the success of KEKB, upgrading to SuperKEKB~\cite{ptep-skekb, SKEKB}, which significantly pushes the luminosity frontier toward more detailed experiments that seek new physics beyond the Standard Model, is considered an urgent issue in elementary particle physics.
The design luminosity of SuperKEKB is $8\times10^{35} \rm{cm^{-2}s^{-1}}$---40 times higher than that achieved by KEKB---and our goal is to accumulate an integrated luminosity of 50 $\rm{ab^{-1}}$.

\begin{figure}[hbt]
\centering
\includegraphics*[width=100mm]{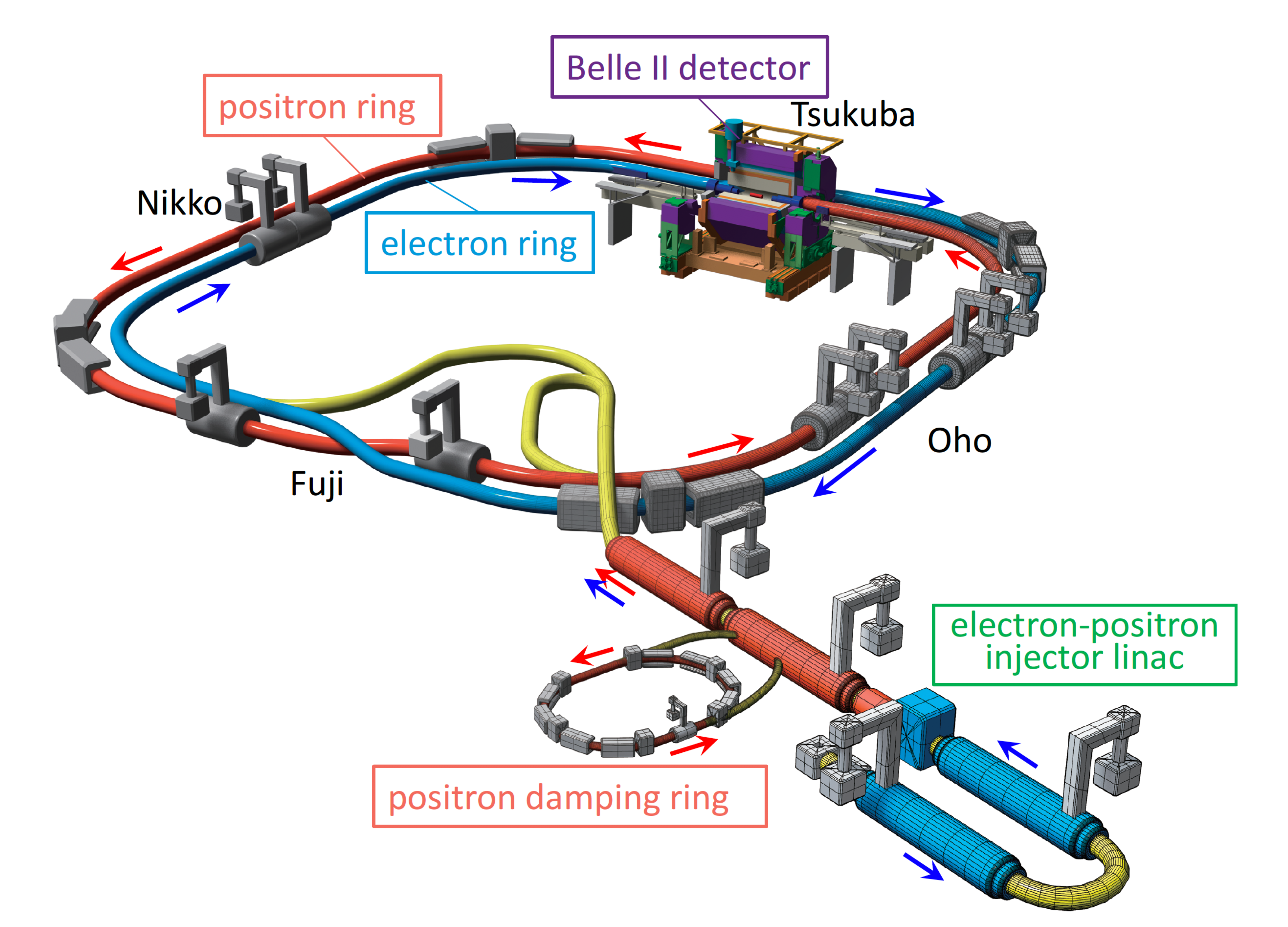}
\caption{Schematic view of SuperKEKB. The electron and positron rings have four straight sections named Tsukuba, Oho, Fuji, and Nikko. The electron and positron beams collide at the interaction point in the Tsukuba straight section.}
\label{superkekb}
\end{figure}

The SuperKEKB collider complex consists of a 7-GeV electron ring (the high-energy ring, HER), a 4-GeV positron ring (the low-energy ring, LER), and an injector linear accelerator (linac) with a 1.1-GeV positron damping ring (DR), as shown in Fig.~\ref{superkekb}. The extremely high luminosity of SuperKEKB required significant upgrades to the HER, LER, and final-focus system of KEKB. The injector linac also required significant upgrades for injection beams with high current and low emittance, as well as for improving simultaneous top-up injections. A new DR was designed and constructed for low-emittance positron beam injection. 
 
Upon approval of the upgrade from KEKB to SuperKEKB, large-scale construction started in 2010.   
After 5.5 years of construction, beam commissioning of SuperKEKB started without the Belle II detector and final-focus superconducting magnets (Phase 1). Phase 1 commissioning was successfully carried out from February to June 2016. After Phase 1, the new final-focus superconducting magnets and the Belle II detector were installed at the interaction region (IR), and renovation of the IR and construction of the DR 
were completed for the start of beam collision tuning  (Phase 2), scheduled for early 2018. This paper reviews the design, construction, and beam commissioning of SuperKEKB. 
 
\section{Overview of the upgrade to SuperKEKB}

\subsection{Collider ring design}
  
Achieving higher luminosity in ring colliders requires higher beam currents $I_{\pm}$, larger vertical beam--beam tune-shift parameters $\xi^*_{y\pm}$, and smaller vertical beta functions at the interaction point (IP) $\beta^*_{y\pm}$. 
Luminosity $L$ is given as follows: 
\begin{eqnarray} 
 L = \frac{\gamma_{\pm}}{2er_e} \left(1+{{\sigma^*_y}\over{\sigma^*_x}}\right)\left(\frac{I_{\pm} \xi_{y\pm}}{\beta^*_{y}}\right) \left(\frac{R_L}{R_{\xi_y}}\right),
\label{lumi}
\end{eqnarray}  
\noindent
\noindent
where $\gamma_\pm$ are the Lorentz factors, $r_e$ the classical electron radius, and $\sigma^*_{x,y}$ the beam sizes at the IP. 
Parameters $R_{L}$ and $R_{\xi_y}$ are correction factors for the geometrical loss due to the hourglass effect and the crossing angle at the IP. 
In this equation, $\sigma^*_{x,y}$ and $\beta^*_{y}$ are assumed to be equal in both rings.

From practical viewpoints, such as those of hardware feasibility and operating costs, it is preferable to increase the beam currents minimally.  
Much higher values of $\xi_y$ than those ever achieved in real colliders are impractical.
Thus, to considerably increase the luminosity of SuperKEKB compared to that of KEKB, we pursued much smaller values of $\beta^*_y$. 
 
\begin{figure}[hbt]
\centering
\includegraphics*[width=100mm]{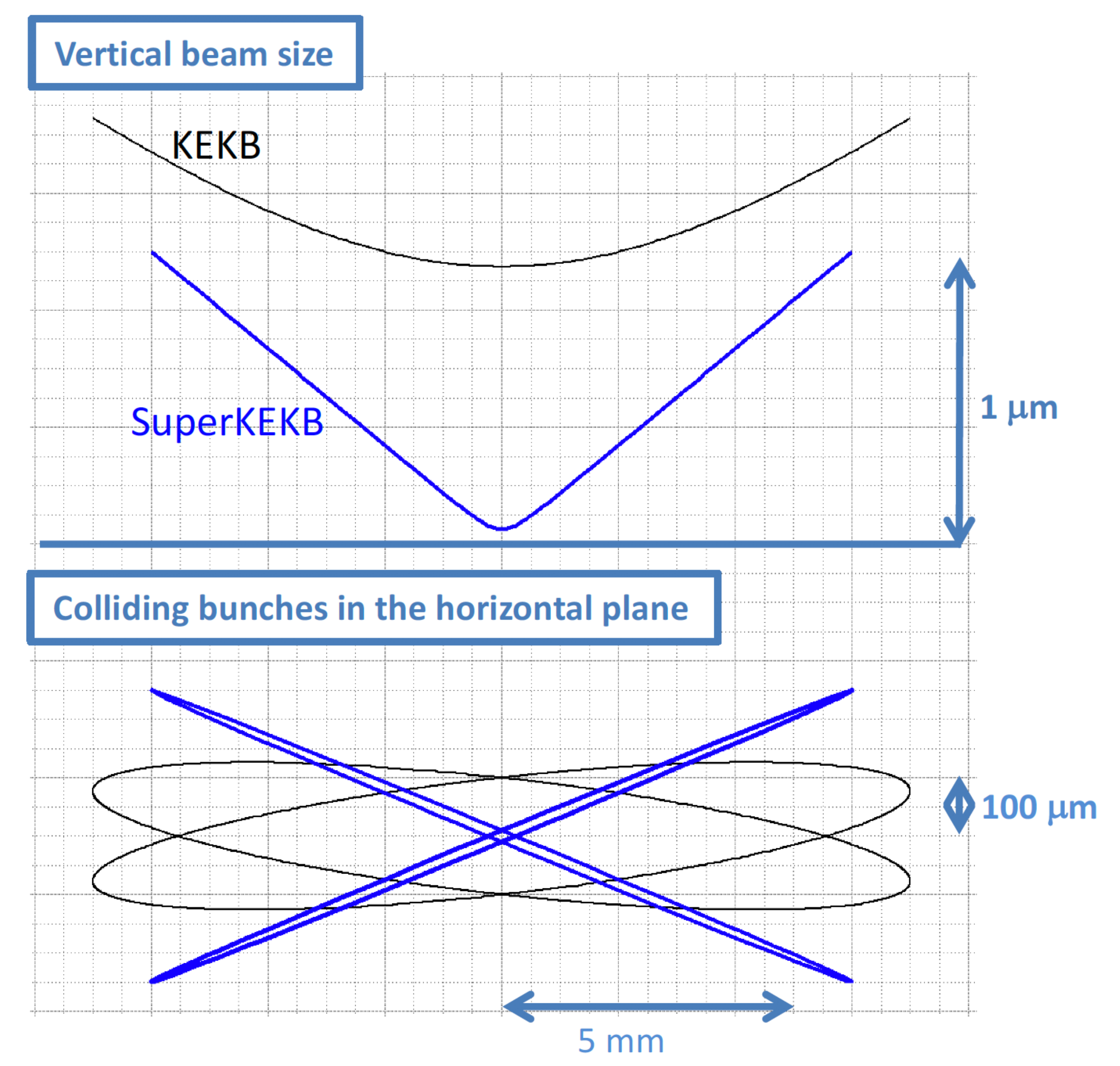}
\caption{Schematic view of the nanobeam collision scheme.}
\label{nanobeam}
\end{figure}

The design strategy for SuperKEKB is based on the nanobeam collision scheme originally proposed by Raimondi~\cite{SuperB}, in which beam bunches with sufficiently small $\sigma^*_x$ collide at a large horizontal crossing angle, as shown in Fig.~\ref{nanobeam}.
In other words, we adopted a large Piwinski angle ($\phi_{\rm{Piw}}\equiv{{\theta_x\sigma_z}/{\sigma^*_x}}\sim20$, where $\theta_x$ is the half horizontal crossing angle).
The longitudinal size of the overlap between colliding bunches decreases by the Piwinski angle as $\sigma_z / \phi_{\rm{Piw}}$, which is much shorter than the bunch length $\sigma_z$.
Therefore, $\beta^*_y$ can be expected to be squeezed to $\sim \sigma_z / \phi_{\rm{Piw}}$, avoiding the hourglass effect.   
To achieve a large $\phi_{\rm{Piw}}$, $\theta_x$ must be sufficiently large, and $\sigma^*_x$ sufficiently small, which means that both low horizontal emittance $\varepsilon_x$ and low  $\beta^*_x$ are required. 

The machine parameters of SuperKEKB and KEKB are listed in Table~\ref{param1}; to summarize this comparison, SuperKEKB's beam currents are doubled, its $\xi_y$ are almost the same as those of KEKB, and its $\beta_y^*$ are reduced by a factor of 1/20.
Thus, we can expect a luminosity 40 times higher than that of KEKB. 
The main features of SuperKEKB are: 

\begin{itemize}
\item \vspace{-0.3cm} Low $\beta^*_y$ of $\sim$300 $\mu$m.
\item \vspace{-0.3cm} Low $\beta^*_x$ of $\sim$30 mm, which is roughly one-tenth of the design value of KEKB (330 mm, although a larger $\beta^*_x$ was used in operation with the crab crossing of KEKB).
\item \vspace{-0.3cm} Low emittances and flat beams:  2$\sim$5 nm (horizontal) and 9$\sim$12 pm (vertical).
\item \vspace{-0.3cm} Large Piwinski angles of $\sim$20.
\item \vspace{-0.3cm} Modest bunch lengths of 5--6 mm.
\item \vspace{-0.3cm} Modest vertical beam--beam parameters of $\sim$0.09. 
\item \vspace{-0.3cm} Short beam lifetimes of $\sim$360 sec. Low $\beta^*_y$ and $\beta^*_x$ decrease the dynamic aperture, resulting in short beam lifetimes and strict requirements for injection beams. Thus, a powerful injector that can deliver low-emittance beams is necessary.
\end{itemize}

The beam energies for SuperKEKB have been changed from those of KEKB --- from 8 to 7 GeV for the HER and from 3.5 to 4 GeV for the LER --- keeping the center-of-mass energy at $\Upsilon$(4S).
The higher energy for the LER was selected to improve Touschek beam lifetime and to suppress emittance growth due to intrabeam scattering. 
The lower energy for the HER decreases the horizontal emittance and synchrotron radiation power. 
Smaller energy asymmetry than that of KEKB is still acceptable for the physics experiment.

\begin{table*}[ttt]
\begin{center}
\caption{Machine Parameters of KEKB and SuperKEKB.  
Values in parentheses for SuperKEKB denote parameters without intrabeam scattering. 
Note that horizontal emittance increases by 30\% owing to intrabeam scattering in the LER.
The KEKB parameters are those achieved at the crab crossing~\cite{ptep-kekb}, where the effective crossing angle was 0. 
(*)Before the crab crossing, the luminosity of $\rm 1.76 \times 10^{34}$$\rm cm^{-2}s^{-1}$ was achieved at the half crossing angle of 11 mrad, where $\phi_{\rm{Piw}}\sim1$~\cite{ptep-kekblattice}. }
\small{
\begin{tabular}{lc|c c|c c|c}

\hline \hline
  &   & \multicolumn{2}{c|}{KEKB}  & \multicolumn{2}{c|}{SuperKEKB} &  \\
  &   & LER (e+)   & HER (e-) &  LER (e+)   & HER (e-) & Units \\
\hline
 
Beam energy & \it{E} & 3.5 & 8.0 & 4.0 & 7.007 &
GeV \\
 
Circumference &
$C$& \multicolumn{2}{c|}{3016.262} & \multicolumn{2}{c|}{3016.315} &
m \\
    
Half crossing angle &
$\theta_x$& \multicolumn{2}{c|}{0 ($11^{(*)}$)} & \multicolumn{2}{c|}{41.5} &
mrad \\

Piwinski angle &
$\phi_{\rm{Piw}}$&
0 &  0 &
24.6 & 19.3 &
rad \\

Horizontal emittance &
$\it \varepsilon_x$ &
18 & 24 &
3.2 (1.9) & 4.6 (4.4) &
nm \\

Vertical emittance &
$\it \varepsilon_y$ &
150 & 150 & 
8.64 & 12.9 &
pm \\

Coupling &&
0.83 & 0.62 &
0.27 & 0.28 &
\% \\

Beta function at IP &
$\beta_x^*/\beta_y^*$ &
1200/5.9 & 1200/5.9 &
32/0.27 & 25/0.30 &
mm \\

Horizontal beam size &
$\sigma_x^*$ &
147 & 170 &
10.1 & 10.7 &
$\rm \mu m$ \\

Vertical beam size &
$\sigma_y^*$ &
940 & 940 &
48 & 62 &
nm \\

Horizontal betatron tune & $\nu_x$ & 45.506 & 44.511 & 44.530 & 45.530 & \\
 
Vertical betatron tune & $\nu_y$ & 43.561 & 41.585 & 46.570 & 43.570 & \\

Momentum compaction &
$\it  \alpha_p$ &
3.3 &  3.4 &
3.20 &  4.55 &
$10^{-4}$\\
 
Energy spread &
$\sigma_\varepsilon$ &
 7.3 & 6.7 &
 7.92(7.53) & 6.37(6.30) &
$10^{-4}$ \\


Beam current &
\it{I} &
1.64 & 1.19 &
3.60 & 2.60 &
A \\

Number of bunches &
$\it n_b$ &
\multicolumn{2}{c|}{1584} &
\multicolumn{2}{c|}{2500} &
 \\
 
Particles/bunch &
\it{N} &
6.47 & 4.72 &
9.04 & 6.53 &
$10^{10}$ \\

Energy loss/turn &
$\it U_0$ &
 1.64 & 3.48 &
1.76 & 2.43 &
MeV \\

Long. damping time &
$\tau_z$ &
21.5 & 23.2 &
22.8 & 29.0 &
msec \\

RF frequency &
$\it f_{RF}$ &
\multicolumn{2}{c|}{508.9} &
\multicolumn{2}{c|}{508.9} &
MHz \\

Total cavity voltage &
$\it V_c$  &
8.0 & 13.0 &
9.4 & 15.0 &
MV \\

Total beam power &
$\it P_b$  &
$\sim$3 & $\sim$4 &
8.3 & 7.5 &
MW \\

Synchrotron tune &
$\nu_s$ &
-0.0246 & -0.0209 &
-0.0245 & -0.0280 &
 \\
 
Bunch length &
$\sigma_z$ &
$\sim$7 & $\sim$7 &
6.0 (4.7) & 5.0 (4.9) &
mm \\

Beam--beam parameter & 
$\xi_x/\xi_y$&
0.127/0.129 & 0.102/0.090 &
0.0028/0.088 & 0.0012/0.081 &
 \\

\hline
Luminosity &
\it{L} &
\multicolumn{2}{c|}{$\rm 2.108 \times 10^{34}$} &
\multicolumn{2}{c|}{$\rm 8 \times 10^{35}$} &
$\rm cm^{-2}s^{-1}$  \\
Integrated luminosity &
\it{$\int L$} &
\multicolumn{2}{c|}{1.041} &
\multicolumn{2}{c|}{50} &
$\rm ab^{-1}$  \\
\hline

\end{tabular}
}
\label{param1}
\end{center}
\end{table*}


\subsection{Major upgrades from KEKB}

To upgrade KEKB to SuperKEKB, new equipment was employed, including products developed to meet the challenging requirements for SuperKEKB, whereas the existing tunnel, infrastructure, and accelerator components for KEKB were reused wherever possible. 
Various existing accelerator components were also improved and modified. Items that received major upgrades are summarized below:

\begin{itemize}

\item 
The KEKB arc sections consisted of 2.5$\pi$ unit cells~\cite{UnitCell,ptep-kekblattice} 
and have wide tuning ranges on both $\varepsilon_x$ and the momentum compaction factor $\alpha_p$.
By fully utilizing their tuning range, requirements for $\varepsilon_x$ and $\alpha_p$ can be realized while nearly preserving the lattice structure of KEKB.
To achieve the design value of $\varepsilon_x$ in the LER, the main dipole magnets were replaced with longer ones,
and the wiggler sections were reformed to have twice as many wiggle pitches by adding new magnets.  
The arc sections in the HER were reused because $\varepsilon_x$ can be decreased acceptably by adjusting quadrupole magnets in the 2.5$\pi$ cells. 
To further reduce $\varepsilon_x$, a new wiggler section was built in the HER, with some of the old LER wiggler magnets reused. 

\item For the new collision scheme with extremely low $\beta_{y}^{*}$, a new final-focus superconducting magnet system (QCS) was employed, which required state-of-the-art design and technology. 
Details of the QCS design are presented in Section~\ref{subsec:QCS}.
The beam lines for the $\sim$300 m final-focus sections are fully reconstructed in both rings. 
In this region, to correct large chromaticity due to small values of $\beta_x^*$ and $\beta_y^*$, 
local chromaticity correction (LCC) sections for both the vertical and horizontal planes were installed in both rings, as shown in Fig.\ref{LCC}.
A pair of identical sextupole magnets were placed in each LCC, connected by the pseudo $-I$ transformation.

\begin{figure}[hbt]
\centering
\includegraphics*[width=100mm]{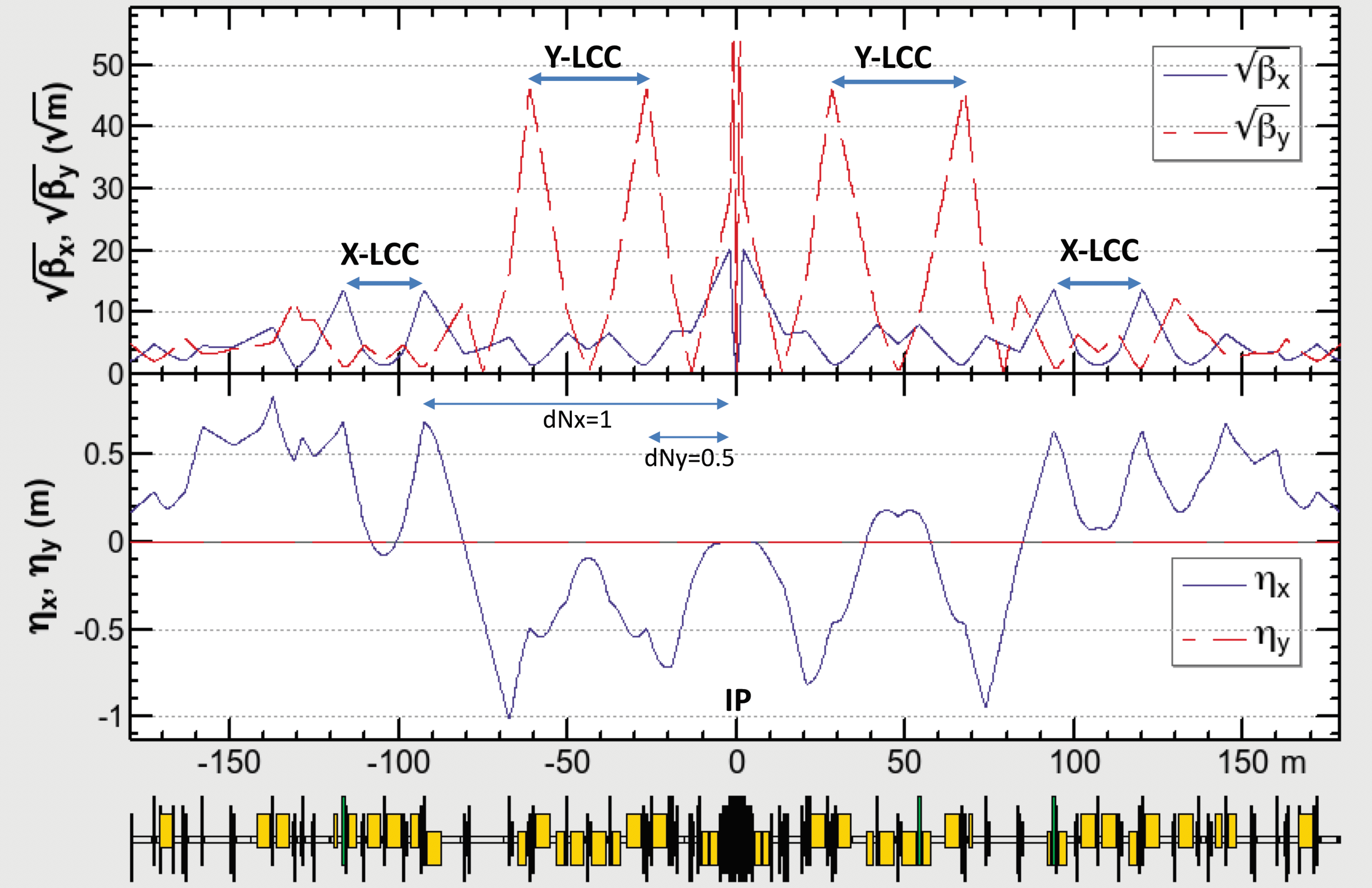}
\caption{Beam optical functions in the final-focus section of the LER.}
\label{LCC}
\end{figure}

\item To cope with the electron cloud effect (ECE) in the LER, more vigorous countermeasures than ever are required for SuperKEKB. 
Based on ECE research that was conducted at KEKB and other machines, various measures are taken, including replacement of beam pipes in most of the ring with new antechamber pipes with an internal TiN coating, groove-shaped surfaces in the dipole magnets, and clearing electrodes in the wiggler magnets and solenoids in field-free regions. 
 
\item To increase beam currents to twice that achieved in KEKB, vacuum components are upgraded for lower impedance and higher thermal strength in addition to the countermeasures to the ECE. The RF system is reinforced to deliver higher power to beam, with normal-conducting cavities composed of three-cavity system called ARES and superconducting cavities (SCCs) reused after necessary improvements to their input couplers and higher-order mode (HOM) dampers. The cooling system for magnets and vacuum components is also reinforced to approximately double its cooling capacity. 
 
\item The beam instrumentation and control systems, including the beam position monitors, beam size monitors, bunch-by-bunch feedback system, and collision feedback system, are upgraded to provide the higher performance required for the lower emittance beams and smaller beam size at the IP.

\item For the required beam quality and injection efficiencies into the LER and HER, the injector linac was upgraded to include a new low-emittance RF electron gun, improvements to the positron source, and the implementation of pulse magnets. A new 1.1~GeV positron DR was built to reduce positron beam emittance before injection into the LER. 

\end{itemize}


\subsection{Final-focus superconducting magnet system}\label{subsec:QCS}

\begin{figure}[hbt]
\centering
\includegraphics*[width=100mm]{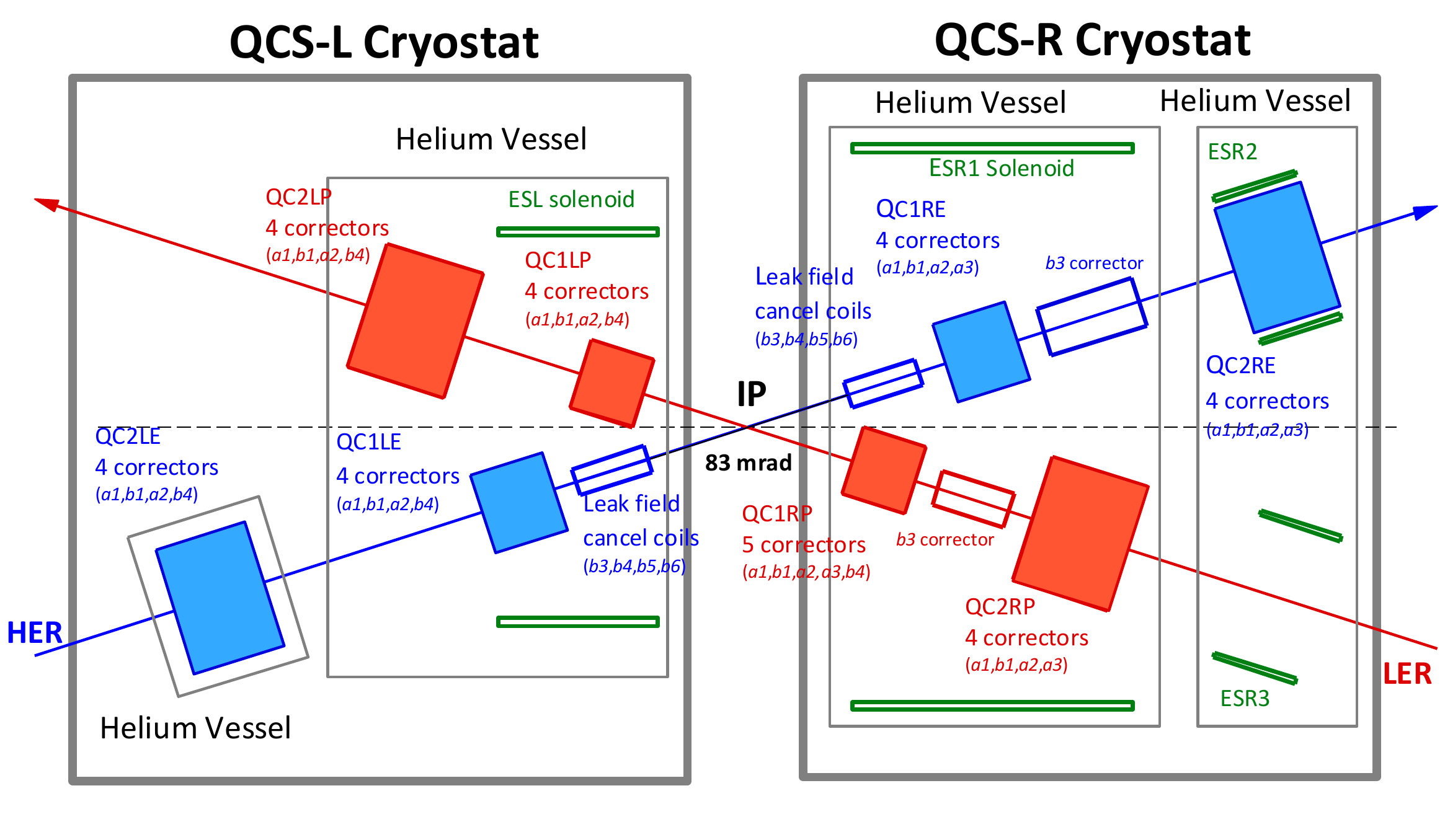}
\caption{Schematic view of the QCS system~\cite{QCSsystem}, which is equipped with various corrector magnets: vertical and horizontal dipole (a1 and b1), skew quadrupole (a2), skew and normal sextupole (a3 and b3), normal octupole, decapole, and dodecapole (b4, b5, and b6) magnets.}
\label{qcs}
\end{figure}

The QCS system, which is a very precise and complex system for realizing extremely small $\beta_y^*$, consists of eight main quadrupole magnets, 43 corrector magnets, and four compensation solenoid coils, as shown in Fig.~\ref{qcs}~\cite{QCSsystem}.
The main quadrupole magnets, QC1's and QC2's, form a doublet for each beam. 
Each quadrupole magnet has four or five corrector magnets, which are used to correct misalignments of the quadrupole magnets, adjust the beam orbit, and optimize the dynamic aperture.
Multipole components of leakage fields from QC1LP and QC1RP that have no yokes are canceled by other corrector magnets on the HER beam line.  
The corrector magnets are fabricated by applying direct winding method of superconducting wire in collaboration with BNL~\cite{QCS_pac13_Parker}. 
The compensation solenoid coils cancel the solenoid field of the Belle II detector so that $\int B_zds =0$ on each side of the IP. 
Field profiles of $B_z$ along the beam lines are carefully optimized to not increase vertical emittance.  
All of the QCS magnets are accommodated in cryostats in the left and right sides of the IP.

\subsection{Requirements for the injector linac and DR}

The injector beam parameters needed to meet the short beam lifetime and small dynamic apertures of both the HER and LER are summarized in Table~\ref{param2}. 
The DR is indispensable for satisfying the requirements for the positron beam.
A ``reverse-bend FODO'' cell was developed to achieve sufficiently short damping times~\cite{DR_design_NIMA, DR_design_ipac10}.
The main DR parameters are listed in Table~\ref{param3}. 
\begin{table}[htb]
\begin{center}
\caption{Injector beam parameters.
}
\small{
\begin{tabular}{lccl}

\hline
Beam &  Positron  &  Electron &  \\ 
Beam energy &  4.0  & 7.007 & GeV \\ 
Normalized emittance $\gamma \it \varepsilon_{x/y}$ & 100/15 & 40/20 & $\mu$m\\
Energy spread & 0.16 & 0.07 &\% \\
Bunch charge & 4 & 4 & nC \\
No. of bunches/pulse & 2 & 2 & \\
Repetition rate & \multicolumn{2}{c}{50} & Hz \\
\hline

\end{tabular}
}
\label{param2}
\end{center}
\end{table}

\begin{table}[htb]
\begin{center}
\caption{Parameters of the positron DR~\cite{SKEKB, DR_design_ipac10}.
}
\small{
\begin{tabular}{lcl}

\hline
Beam energy &  1.1  & GeV \\ 
Circumference & 135.498 & m \\  
Number of bunch trains & 2 & \\  
Number of bunches / train & 2 & \\  
Damping time ($\tau_x$/$\tau_y$/$\tau_z$) & 11.5/11.7/5.8  & ms \\
Injected-beam emittance & 1400 & nm \\
Equilibrium emittance ($\it \varepsilon_x$/$\it \varepsilon_y$) & 41.5/2.08 & nm\\
Emittance at extraction ($\it \varepsilon_x$/$\it \varepsilon_y$) & 42.9/3.61 & nm \\
Momentum compaction & 0.0141& \\
Energy spread & 5.5 $\times 10^{-4}$ & \\
Energy loss/turn & 0.0847& MeV \\
Total cavity voltage & 1.44 & MV \\
RF frequency & 508.9 & MHz \\
Bunch length & 6.5 & mm \\
\hline

\end{tabular}
}
\label{param3}
\end{center}
\end{table}



\section{Construction and commissioning scenario}

Discussions between the accelerator and Belle II groups led to the adoption of a three-phase beam commissioning scenario~\cite{MR_all_ipac13, MR_all_ipac14}. The phased commissioning plan minimizes the risk of damage to the Belle II detector from insufficient beam conditions or accidents before stable operation conditions are established. For example, the Belle II group required sufficient vacuum scrubbing of the LER and HER before installation of Belle II, with a beam dose up to several hundred A$\cdot$hour, to sufficiently reduce beam background. Consequently, the main tasks of Phase 1 commissioning, which was performed without QCS and Belle II, were basic machine tuning, low-emittance beam tuning, and vacuum scrubbing with beam currents up to about 0.5--1.0~A. After installation of QCS and Belle II, Phase 2 commissioning 
is performed. However, the vertex detectors (VXD) 
at the 
center of Belle II 
are not installed in this phase. Beam collision tuning within the nanobeam collision scheme will be performed by gradually squeezing $\beta_{y}^{*}$. After sufficient collision beam performance and understanding of the radiation to Belle II are confirmed, the VXD will be installed, and Phase 3 operation with full Belle II detectors will start. Beam tuning will continue to increase the luminosity with gradually increasing beam currents.

\begin{figure}[hbt]
\centering
\includegraphics*[width=100mm]{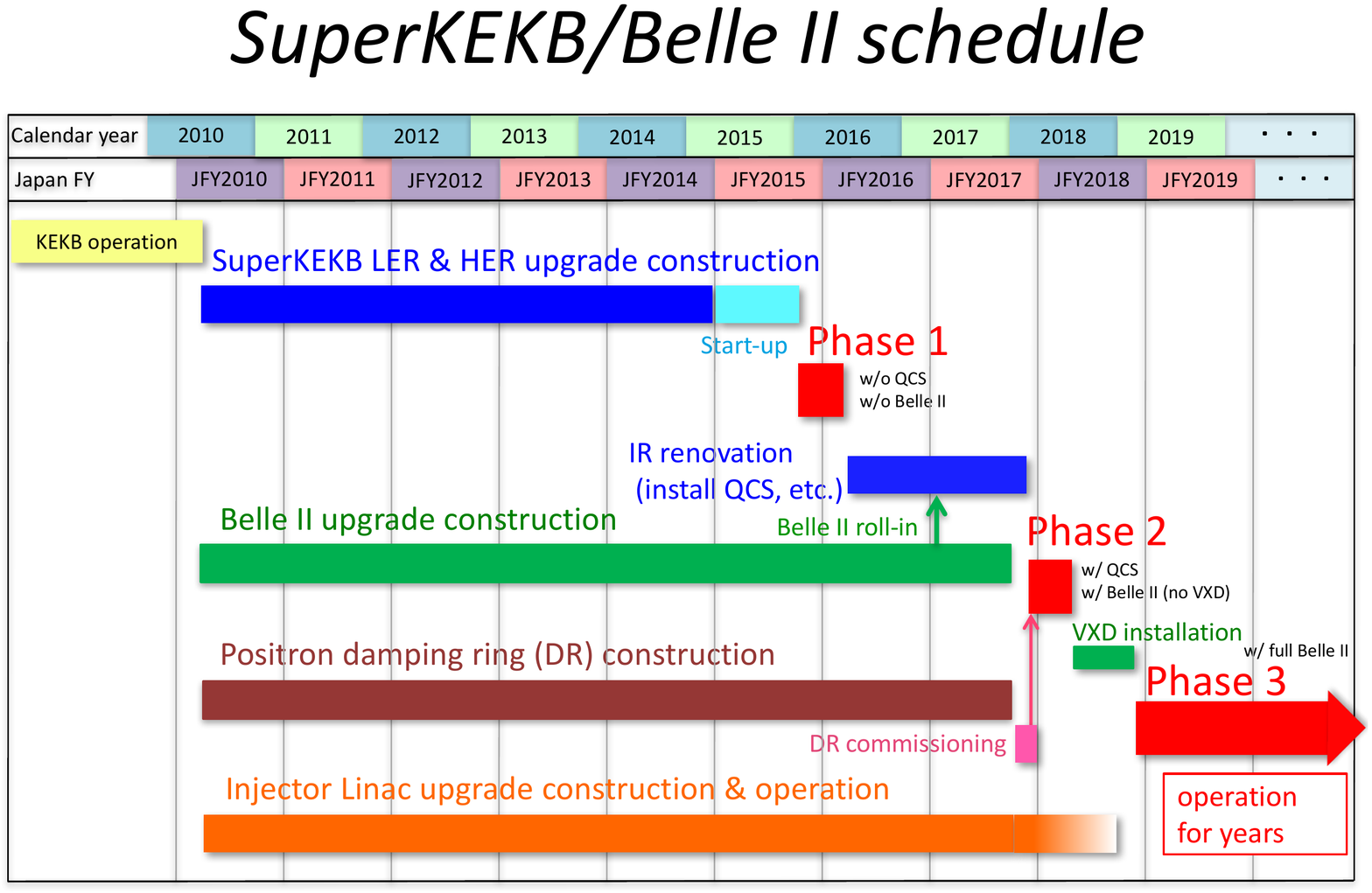}
\caption{Overall schedule of SuperKEKB/Belle II project.}
\label{Fig:Schedule}
\end{figure}

The construction and startup work strategies were optimized to meet the phased commissioning scenario, as well as various boundary conditions such as budgetary profiles, design progress, and technical issues. The overall schedule of the SuperKEKB/Belle II project is shown in Fig.~\ref{Fig:Schedule}. Construction and startup work for upgrading the LER and HER for Phase 1 operation was completed in January 2016. Most of the major upgrades to the vacuum, magnet, and beam instrumentation systems for the LER and HER were completed by this date. With the RF system reinforced before Phase 1, about 70\% of the design beam currents in both rings can be stored, which is sufficient for operation in Phases 1 and 2. Further reinforcement of the RF system is foreseen during Phase 3, depending on beam condition requirements and budgeting. 

The phased scenario allowed sufficient time for the design, development, and fabrication of several new components needed for Phase 2, (e.g., the QCS system, which brought very complicated design and technical challenges). 
Fabrication of the QCS for the left side of the IP (QCSL) started in 2012 and was completed in 2015.
Fabrication of the QCS for the right side of the IP (QCSR) started in 2013 and was completed in February 2017. Field measurements of the QCSL and QCSR in the beam line were completed in August 2017. 
As another example, after we confirmed the performance of two prototype antechamber-type collimators in Phase 1 with high current beams, mass production of the collimators needed for Phase 2 began. In the phased scenario, the DR is not needed for Phase 1, but is indispensable for Phase 2. The DR construction and startup schedule was then optimized to start beam commissioning prior to the start of the main ring (MR; LER and HER) Phase 2, so that DR beam tuning can be performed well enough to inject beams into the LER via DR when the MR is ready for Phase 2.

\section{Main ring hardware upgrade}

\subsection{Magnet system}

\begin{figure}[hbt]
\centering
\includegraphics*[width=80mm]{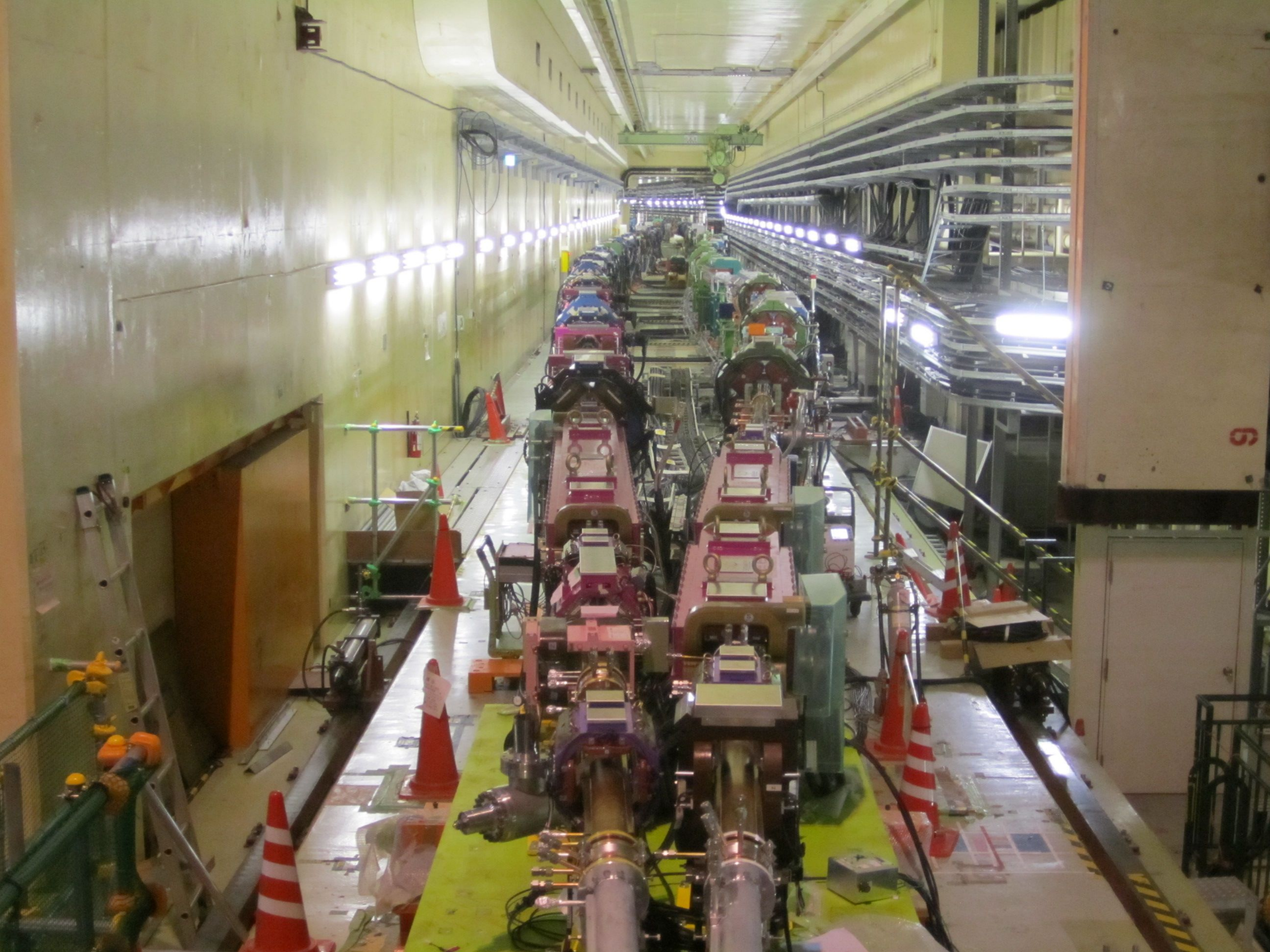}
\caption{New beam lines in the Tsukuba straight section.}
\label{Fig:Tsukuba magnets}
\end{figure}

For the new optics design, many magnets and power supplies for the magnets must be replaced, rearranged, and added, although the present cell structures in the arc sections are basically preserved in both rings. The 100 old 0.89-m-long main dipole magnets in the LER arc sections were replaced with new 4.2-m-long magnets. The LER wiggler sections in the Oho and Nikko straight sections were rearranged to have twice as many wiggle pitches by adding new half-pole and single-pole wiggler magnets to the existing double-pole magnets. In the HER, a new wiggler section was formed using recycled wiggler magnets taken from the LER. In the Tsukuba straight section (both sides of the IP, each about 150 m in length), in which local chromaticity correction sections for both the vertical and horizontal planes were implemented, new beam lines for both rings were built with newly fabricated magnets, as shown in Fig.~\ref{Fig:Tsukuba magnets}. In the arc sections, the main dipole magnets in the HER and most of the quadrupole magnets in the LER and HER have been reused. 

To improve optics control around the IR, an innovative idea was proposed: tilting sextupole magnets with variable angles to control the ratio of the skew sextupole field component to the normal sextupole field component. A new table, on which to mount a recycling sextupole magnet, was developed to allow tilting of the sextupole magnet from -30$^{\circ}$ to +30$^{\circ}$, with a high setting accuracy of 0.1~mrad. Twenty-four sets of sextupole magnets mounted on a tilting table were fabricated and installed~\cite{MR_mag_ieeesc26_TiltSextupole}. 

Power supplies for the QCS magnets must have high current stability, low current ripple, high current setting resolution, and a quench protection system. Power supplies for the QCS main magnets (DC 2~kA, 10~V) and corrector magnets (DC $\pm$60~A, $\pm$5~V) were developed to meet these requirements~\cite{MR_mag_ieeesc26_MagPs}. MW-class power supplies for bending and wiggler magnets were also newly fabricated to replace old ones. More than 2000 existing large (0.1--0.5~MW), medium ($\le$100~kW), and small ($\le$3~kW) power supplies from KEKB are 
reused after necessary overhaul, repair, and modification~\cite{MR_mag_ipac16_MagPs}.

Owing to the Great East Japan Earthquake, not only did the alignment of existing magnets go out of position, but the reference points of the magnet alignment system in the tunnel became useless. The alignment targets were rebuilt, and all existing and newly installed magnets were aligned. The high accuracy of the alignment was qualified in that the difference in the circumference between the two rings found in Phase 1 operation was only 0.2~mm, well adjustable within the range of the chicane magnets of the LER~\cite{MR_mag_ipac16_MagAll}.

\subsection{Vacuum system} 

The requirements for upgrading the vacuum system come from higher beam currents in both rings, as well as the ECE in the LER. (1) The synchrotron radiation (SR) power and photon density are high, especially in the wiggler sections. (2) The beam impedance of various vacuum components must be minimized to suppress the excitation of HOM. (3) Effective countermeasures are required to reduce the electron density in the beam pipes~\cite{MR_vac_vacuum121_VacConst}. 

Antechamber beam pipes have been adopted to reduce SR power density at the beam pipe walls and beam impedance, in addition to mitigate the ECE. Between the beam pipes, bellows-chambers with a comb-type RF shield structure are used for high thermal strength. For a stepless connection between these components, we applied Matsumoto--Ohtsuka-type (MO-type) flanges for the first time at a large scale~\cite{MR_vac_jvstA2005_MOflange}. The bellows chambers, MO-type flanges, and gate valves have cross-sections of the same antechamber shape, which minimizes impedance and heating due to beam-induced HOM.

\begin{figure}[hbt]
\centering
\includegraphics*[width=80mm]{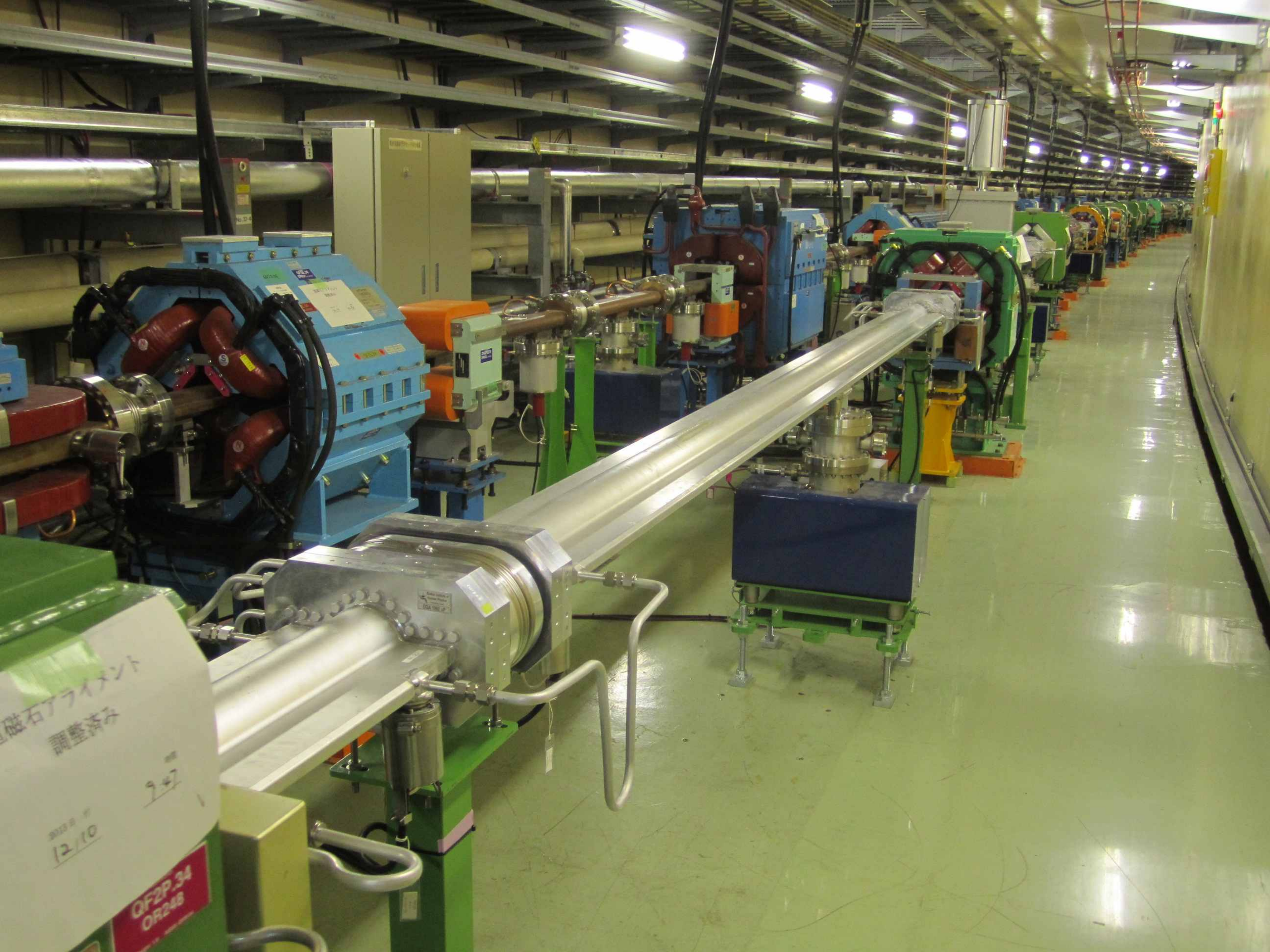}
\caption{Antechamber beam pipes in the LER arc section.}
\label{Fig:Arc vacuum}
\end{figure}

Aluminum beam pipes are easier to fabricate and have lower production costs than copper pipes. We confirmed in beam tests conducted at KEKB that an aluminum surface, if coated with TiN, produces as low a secondary electron yield as that produced by a TiN-coated copper surface, which suppresses the electron density increase due to multipactoring~\cite{MR_vac_jvstA34_VacConst}. The electron cloud density is approximately the same for both materials with TiN coating. Consequently, aluminum coated with a TiN film was adopted for new antechamber beam pipes in the LER arc sections. The TiN coating was performed in a new facility launched at KEK for coating and baking more than 1000 beam pipes. In the LER, beam pipes and related vacuum components in the arc sections and Tsukuba straight section (more than 90\% of the ring) were replaced with new antechamber components. In both rings, beam pipes in the wiggler section, where SR power is high, and the Tsukuba straight section, where the contribution to Belle II background is significant, were replaced with new antechamber-type beam pipes made of copper. The beam pipes in the arc sections of the HER used in KEKB were reused. Figure~\ref{Fig:Arc vacuum} shows the new antechamber beam pipes in the LER arc section.  

In addition to adopting antechamber beam pipes, we have taken various countermeasures to address the ECE. For example, grooved surfaces and clearing electrodes were applied to dipole field regions. The beam pipes in bending magnets in the arc sections were machined with grooves in the upper and lower inner surfaces. A method of forming a thin electrode on the inner surface using a thermal spray developed at KEK was applied to beam pipes in wiggler magnets~\cite{MR_vac_vacuum121_VacConst}. These measures were completed before Phase 1 commissioning.

A new collimator for antechamber beam pipes was developed based on those used in PEP II at SLAC. Two new collimators were installed in the LER and checked with a 1~A beam in Phase 1 commissioning~\cite{MR_vac_prst19_VacCommis}. Expected performance was confirmed, including low temperature rise in the bellows behind the moving collimator head and collimator head positioning accuracy; mass production 
was conducted to add more collimators for Phase 2.

\subsection{RF system} 

One serious concern for high-current, large-circumference storage rings is the excitation of fast-growing longitudinal coupled-bunch instability (CBI) caused by the accelerating mode of RF cavities. Another is minimizing the impedance of HOMs of cavities to avoid longitudinal and transverse CBI. To overcome these difficulties, two types of innovative heavily HOM-damped cavities with large stored energy were utilized in KEKB; one is the ARES cavities, and the other is the SCCs. These cavities performed excellently with high-current beams during KEKB operation~\cite{MR_rf_ptep2013_kekb}. The RF systems used with the ARES cavities and SCCs in KEKB are 
reused as much as possible, with necessary reinforcements and improvements required to meet SuperKEKB beam parameters. In SuperKEKB, the design beam currents in both rings are double those achieved in KEKB and more than twice the power needs to be delivered to the beam, whereas the required RF voltage is about the same as that of KEKB.

\begin{figure}[hbt]
\centering
\includegraphics*[width=100mm]{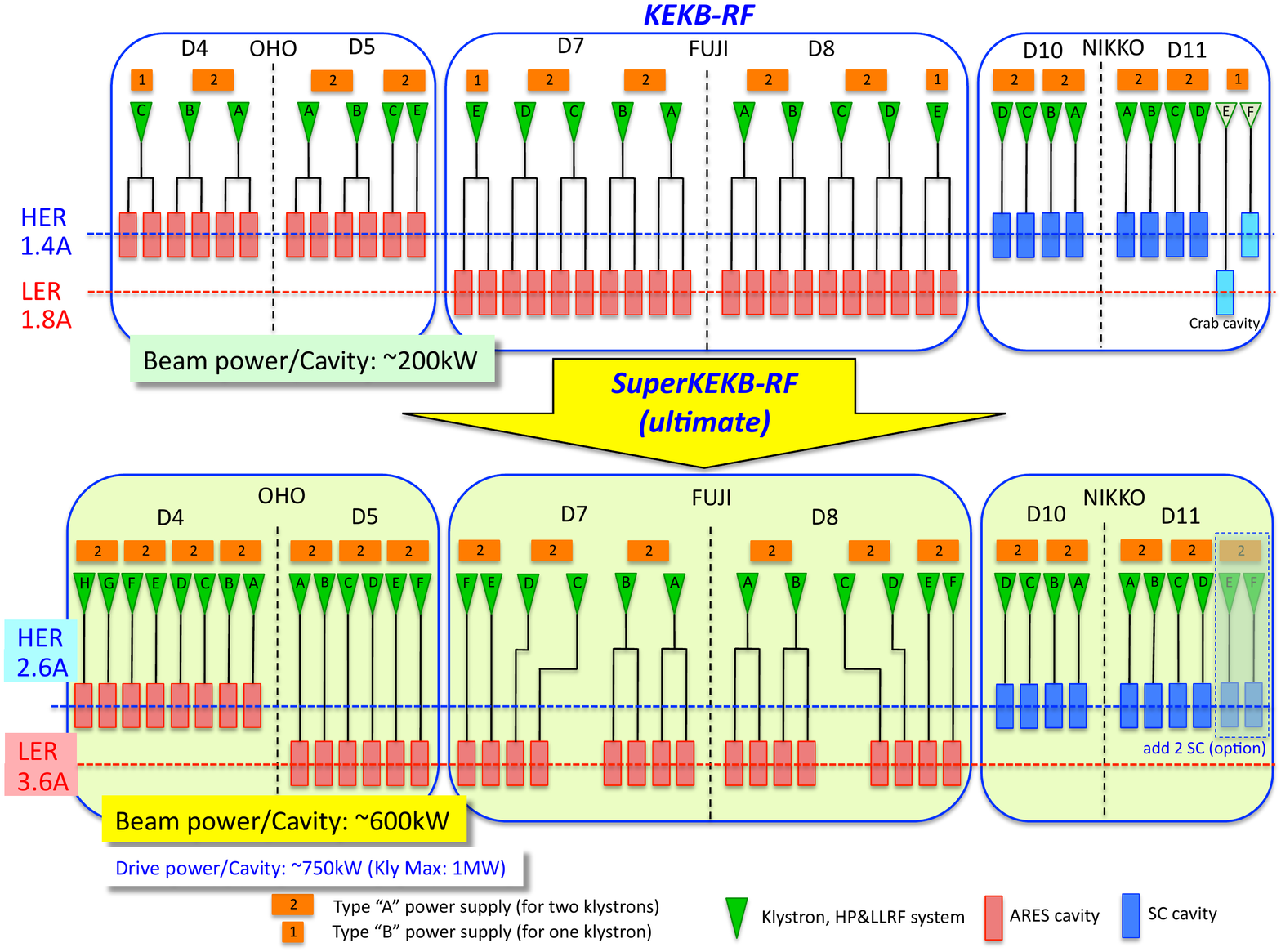}
\caption{RF station configuration at KEKB and SuperKEKB for the design beam current.}
\label{Fig:RF reinforcement}
\end{figure}

To increase the power delivered to the beam per cavity, the system configuration was changed such that one ARES cavity is powered by one klystron; in the previous configuration, two ARES cavities were powered by one klystron. Reinforcement is carried out in two steps: in the first step, which was completed by Phase 1 commissioning, 14 ARES cavities out of 30 were converted to the 1:1 configuration by adding and relocating the cavities, klystrons, and other RF components. The reinforced system can support at least 70\% of the design beam current in both rings. For the next step, four klystrons will be added to convert the remaining eight ARES cavities to the 1:1 configuration.
Figure~\ref{Fig:RF reinforcement} shows the RF station configuration for the design beam current.  

In cavities with the 1:1 configuration, the input couplers of the ARES cavities were upgraded to increase their power handling capability from 400~kW to over~700 kW; this was achieved by increasing the input coupling factor from 3 to 6 by extending the coupling loop from 17~mm to 60~mm in height~\cite{MR_rf_pasj14_InputCoupler}. Another improvement for the input coupler is the adoption of fine grooving on the outer conductor surface of the coaxial line to suppress multipactoring discharge~\cite{MR_rf_prstab13_CouplerMP}.

The present ferrite HOM dampers of the SCC are considered marginal at around 2.0~A in the HER, owing to the increase in beam-induced HOM power with higher beam currents. To achieve the design beam current of 2.6~A in the HER, additional HOM dampers made of SiC are planned to be installed between adjacent cavities, which effectively reduces the HOM loads to the ferrite dampers~\cite{MR_rf_SCC_SiC}. One set of SiC dampers has been installed to check this scheme with the beam in Phase 2 operation. 

The Q-factors at the high fields of several SC cavities have significantly degraded during the long-term operation of KEKB. Although these cavities can still be operated at a nominal operating voltage of 1.5~MV per cavity, this may become an issue for the future long-term operation of SuperKEKB, with its much higher beam current. To recover the cavity performance, a horizontal high-pressure water rinsing (HHPR) system was developed, which can be applied to a cavity without disassembling it from the cryostat. HHPR was applied to a degraded spare cavity, and its performance was successfully recovered. This cavity was then installed in the HER to replace another degraded cavity during operation. The degraded cavity was taken out of the ring and subjected to HHPR. After HHPR, its Q-factor recovered drastically~\cite{MR_rf_SCC_HHPR}. 

A new low-level RF (LLRF) control system, composed of $\mu$TCA-platform FPGA boards with embedded experimental physics and industrial control system (EPICS) input-output controllers (IOC), was developed for higher accuracy and flexibility. In nine RF stations out of 22 for the ARES cavities, the new LLRF system was installed to replace the existing system. These LLRF systems exhibited good performance in Phase 1~\cite{MR_rf_ipac17_LLRF}. 

Because the bunch-gap transient effect may be a more serious concern with higher beam currents and smaller values of $\beta_{y}^{*}$ at the IP, an advanced simulation study was conducted, which investigated transient loading in the ARES three-cavity system. This study clarified that the rapid change in beam phase observed in KEKB operation is caused by the 0 and $\pi$ modes of the ARES, and discussed possible measures for SuperKEKB operation~\cite{MR_rf_prst19_gaptr}.

\subsection{Beam instrumentation and control system}

In the LER, the narrow-band detectors at 1~GHz for beam position monitors (BPMs) were replaced with new ones at 509~MHz, because the cut-off frequency of the new antechamber is lower than 1~GHz. The button electrodes were also changed to new ones, with a reduced diameter of 6~mm (compared to 12~mm for KEKB) for higher beam current~\cite{MR_mon_ibic12_BeamInstrumentation}. For the reused beam pipes in the HER arc sections, the BPM chambers and 1~GHz narrow-band detectors are also reused. New gated turn-by-turn detectors were developed and installed for BPMs at selected positions in both rings~\cite{MR_mon_ibic13_GTBT}. 

Bunch-by-bunch feedback systems for SuperKEKB have been developed in collaboration with SLAC and INFN. The transverse bunch feedback system in the LER and HER consists of position monitors, high-speed signal processing using iGP12 digital feedback filters, strip line kickers, and wide-band high-power amplifiers. A longitudinal feedback system with overdamped (DA$\Phi$NE) kickers~\cite{MR_mon_pasj16_FB} was implemented in the LER. Related instrumentation, including a bunch current monitor (BCM) and a bunch oscillation recorder (BOR), has also been installed~\cite{MR_mon_ibic12_BOR}. 

X-ray beam-size monitors based on coded aperture imaging were developed for high-resolution, bunch-by-bunch measurement of small beam sizes. To support a coded aperture mask pattern in gold on a surface, a diamond substrate was chosen, which was shown in a high-power test at CesrTA to be more robust than a silicon substrate~\cite{MR_mon_ibic13_Xray}. The X-ray monitors were installed in the LER and HER for Phase 1, primarily for vertical beam size measurement. 

Other beam monitoring systems installed before Phase 1 include SR monitors primarily for horizontal beam-size measurement, streak cameras for bunch-length measurement, a large-angle beamstrahlung monitor (LABM)~\cite{MR_mon_PhysRev_LABM} for measurement of polarization components of radiation emitted by beam--beam collisions, and beam loss monitors. 

A collision feedback system~\cite{MR_mon_IPFB} to maintain the optimum beam collision condition 
was prepared for Phase 2. The orbit feedback in the vertical direction will be conducted based on the beam orbit measurement related to beam--beam deflection. This system is similar to that of KEKB, although much faster feedback is required for SuperKEKB. In the horizontal direction, where the beam--beam parameter is small because of the effective horizontal beam size for the nanobeam collision scheme, a dithering system, similar to that of PEP-II, 
has been adopted. The dithering system 
was prepared in collaboration with SLAC.

New control modules have been developed for the SuperKEKB control system, including 
upgraded version of power supply interface controller module (PSICM) for magnet power supplies~\cite{MR_cont_icalepcs15_magcont}, 
EPICS embedded CPU module for programmable logic controller (PLC)~\cite{MR_cont_icalepcs13_PLC}, and
new event modules to deliver triggers to the beam monitors at the DR~\cite{MR_cont_icalepcs15_TimingDR}. 
Beam abort system 
has been upgraded to protect the extract window against higher beam current and smaller beam size~\cite{MR_cont_ipac14_AbortSystem}. The abort trigger system was also improved to achieve faster response time less than 20~$\mu$s~\cite{MR_cont_icalepcs15_AbortTrigger}, compared to $\sim$100~$\mu$s in KEKB.


\section{Damping ring construction}

\begin{figure}[hbt]
\centering
\includegraphics*[width=80mm]{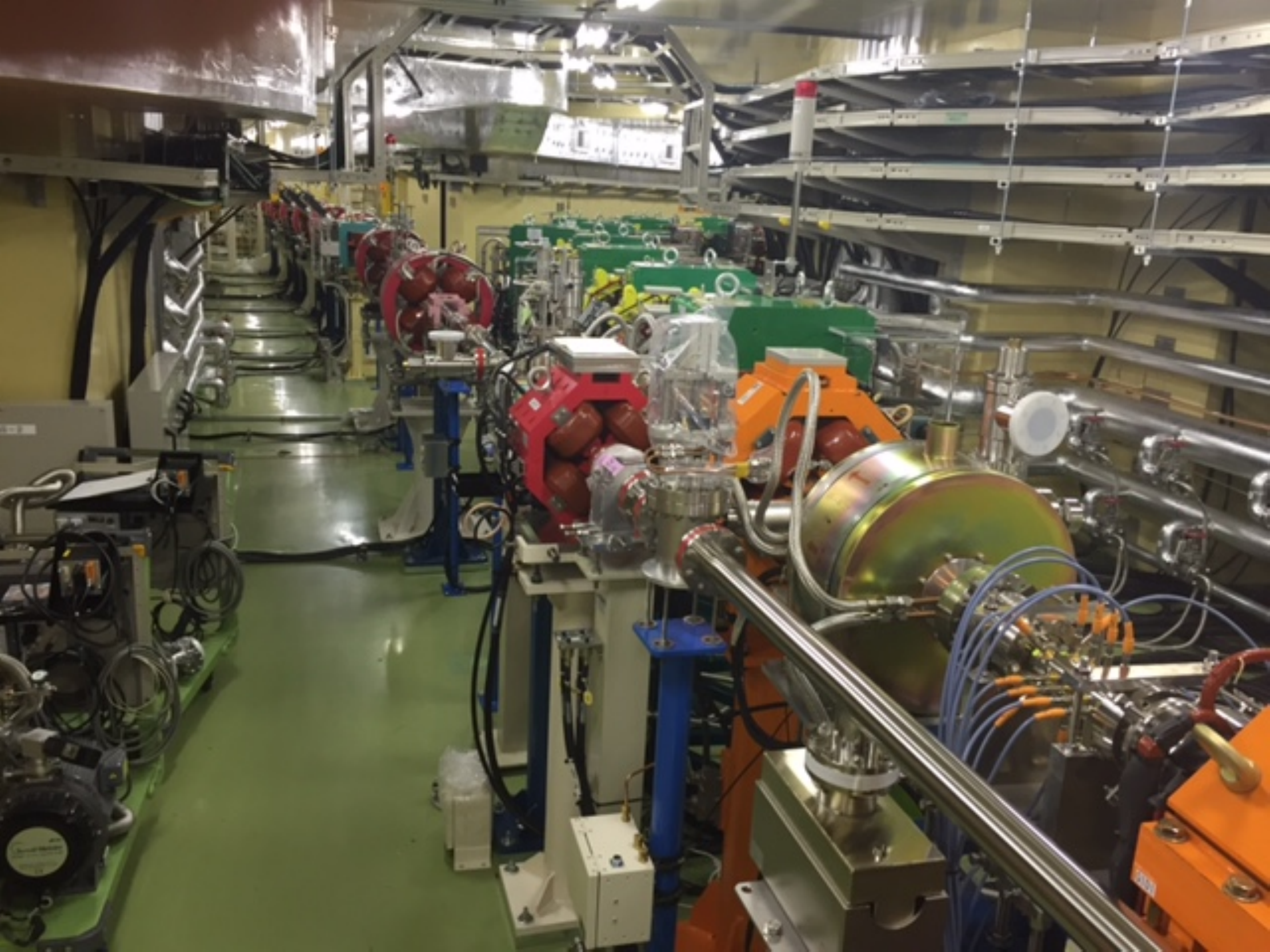}
\caption{DR and injection transport line from the linac.}
\label{Fig:DR and LTR}
\end{figure}

In 2014, a new DR tunnel and two buildings for power supply and facilities were completed, and installation of the accelerator components began. 
Installation of all accelerator components and startup operation were completed, and the DR beam commissioning will start in early 2018 prior to MR Phase 2. Figure~\ref{Fig:DR and LTR} shows the DR arc section and the injection transport line from the linac.  

In the arc sections, antechamber beam pipes made of aluminum with internal TiN coating were adopted as a countermeasure for the ECE. The top and bottom of the inner surface are also grooved. For the reverse-bend FODO lattice~\cite{DR_design_ipac10} two types of bending magnets, which are excited in the opposite polarization, are used. The beam pipes are designed for this configuration; in particular, the SR is radiated on both sides of the pipes~\cite{DR_vac_pasj16}. 

\begin{figure}[hbt]
\centering
\includegraphics*[width=80mm]{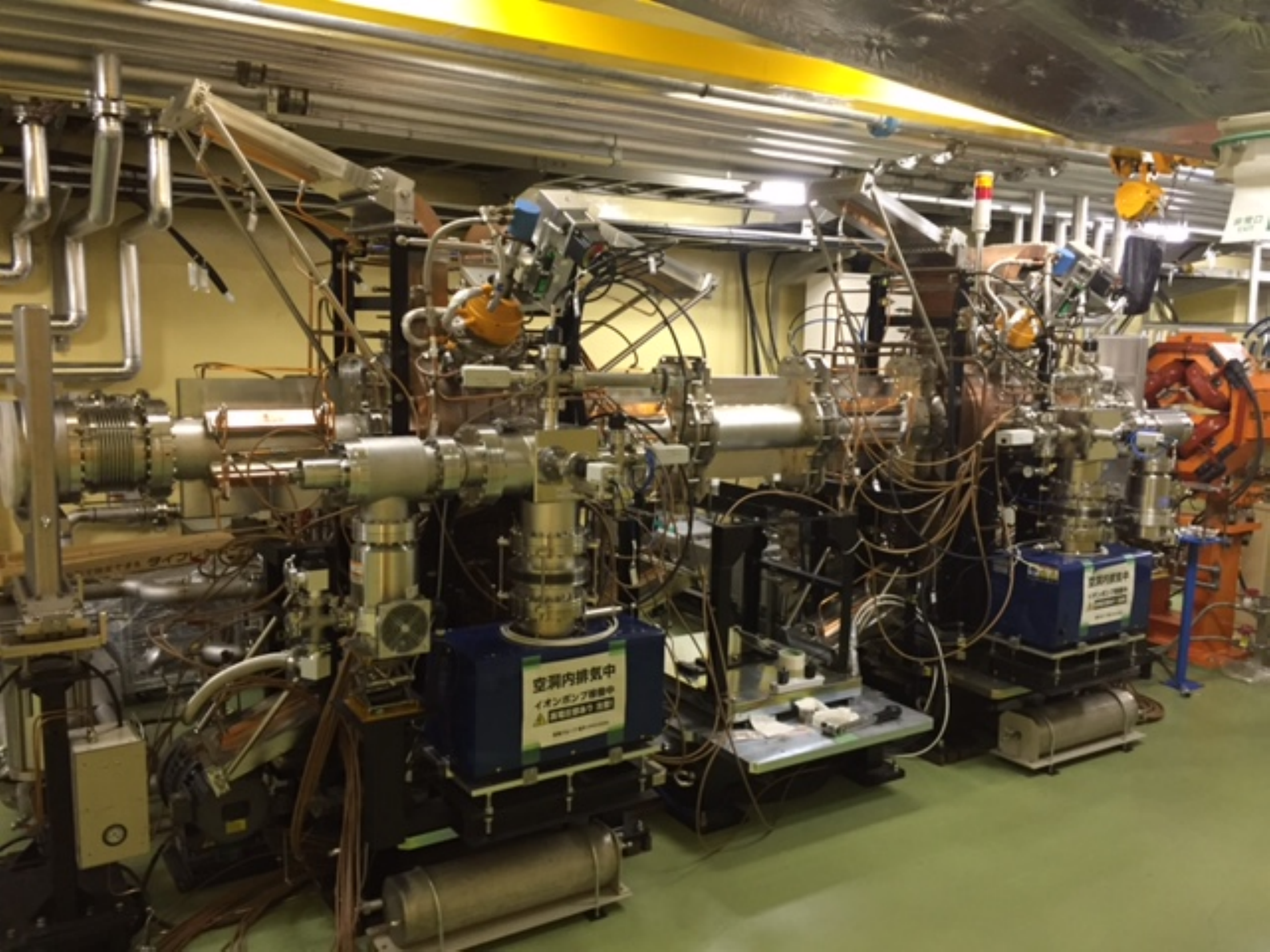}
\caption{Accelerating cavities in the DR.}
\label{Fig:DR cavity}
\end{figure}

A DR accelerating cavity has been newly developed based on the accelerating cavity of the ARES. Because the longitudinal coupled-bunch instability due to the accelerating mode is of no concern in the case of the DR, the storage cavity of the ARES is not needed. 
The required RF voltage of 1.4~MV can be supplied by two cavities operating at 0.7~MV per cavity. Two cavities have been installed in the DR, as shown in Fig.~\ref{Fig:DR cavity}, and high-power operation was successfully demonstrated~\cite{DR_rf_pasj14}. There is a room in the RF section for a third cavity to be installed, if necessary. A klystron, high-power RF system, and new digital LLRF that are similar to those used in the MR have been set up for the DR. 

Beam position monitors, a transverse bunch-by-bunch feedback system, a SR monitor, and loss monitors similar to those used in the MR have been installed for the DR beam instrumentation system. Because of the narrow space for installing button electrodes in beam pipes, the button electrodes were made with a new design, in which two button electrodes are attached to one flange~\cite{DR_mon_ibic13}. 



\section{Injector linac}

The electron--positron injector linac at KEK has delivered electrons and positrons for particle physics and photon science experiments since 1982.  
It was originally constructed for Photon Factory as a 2.5-GeV electron linac beginning in 1978 and was commissioned in 1982.  A positron generator was added for the TRISTAN electron/positron collider project, and was operated until 1995. Then, the linac was upgraded for the KEKB asymmetric-energy collider project with energy enforcement up to 8 GeV~\cite{nim-linac,ptep-linac}.  

Since 2010, KEKB has been further rejuvenated for the SuperKEKB collider project. This project aims at a 40-fold increase in luminosity over the previous KEKB project, 
in order to increase our understanding beyond the standard model of elementary particle physics~\cite{ptep-skekb} after the KEKB project's decade of successful operation. The SuperKEKB asymmetric-energy electron--positron collider, with its extremely high luminosity, requires injection beams with high current and low emittance in the transverse and longitudinal directions~\cite{satoh-ipac16}. It should also perform simultaneous top-up injections into four storage rings and a DR by pulse-to-pulse modulations (PPMs) to avoid interfere between three facilities: SuperKEKB, Photon Factory (PF), and PF Advanced Ring (PF-AR).

\begin{figure}[hbt]
\centering
\includegraphics*[width=120mm]{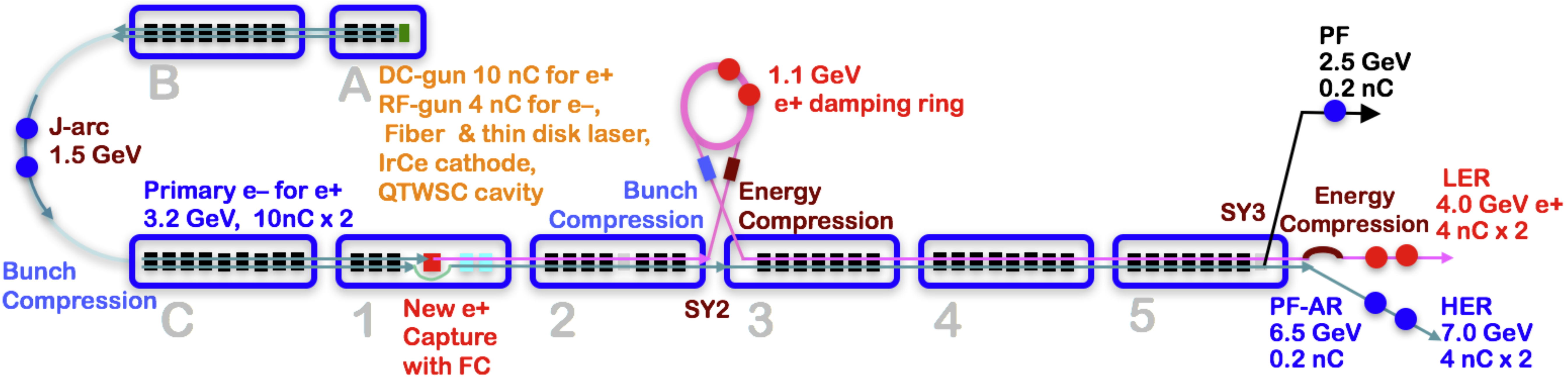}
\caption{Overview of 600-m electron/positron injector linac with 60 accelerating units.}
\label{overview}
\end{figure}

\begin{figure}[hbt]
\centering
\includegraphics*[width=80mm]{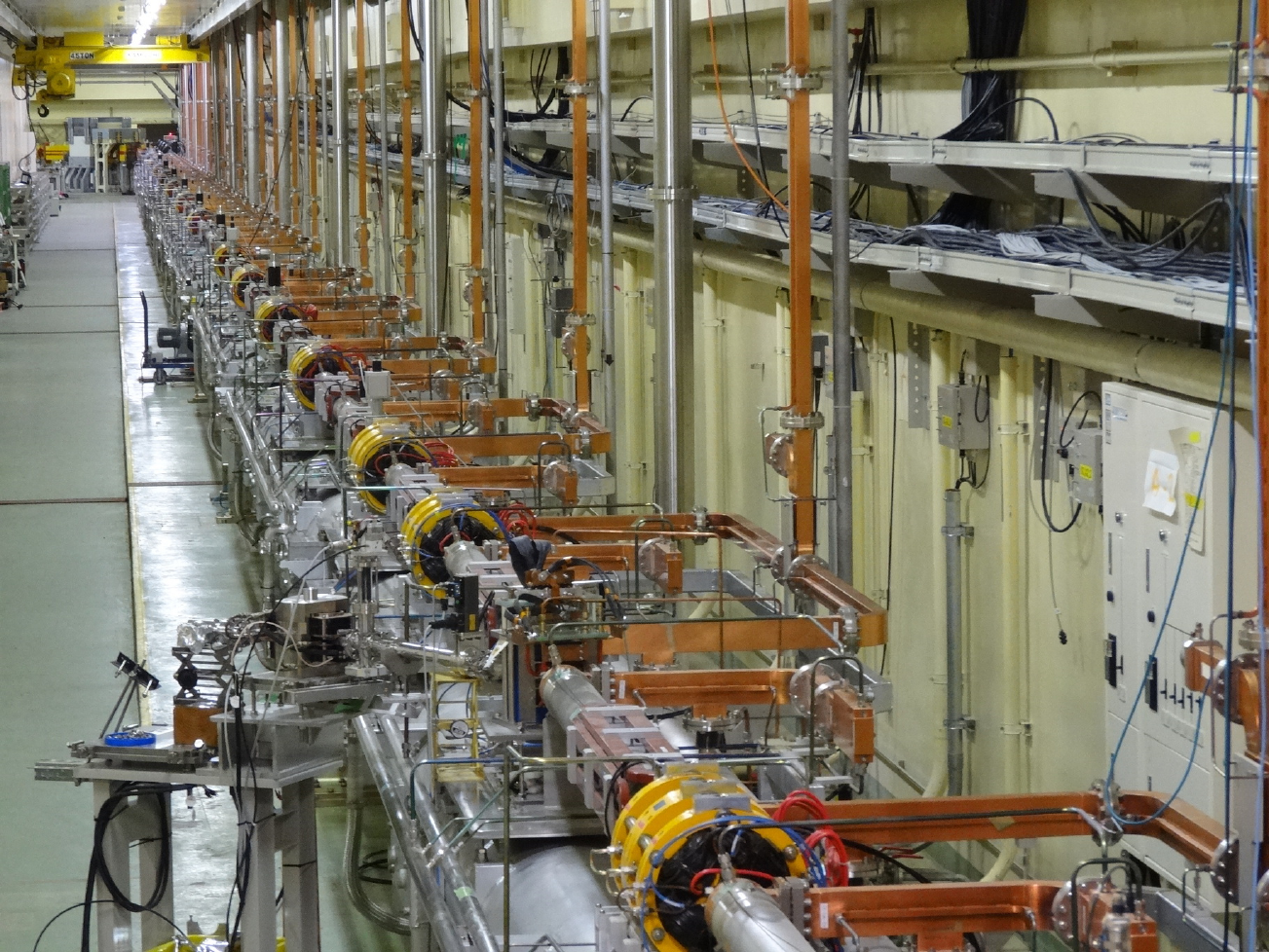}
\caption{A typical part of the injector linac at the A and B sectors.}
\label{absector}
\end{figure}

The 600-m injector linac is composed of 60 high-power accelerating units 
followed by a beam switch yard. 
The overview of the linac is shown in Fig.~\ref{overview}, and the photo of A and B sectors is shown in Fig.~\ref{absector} as a typical injector section.
The injector should meet the requirements of the SuperKEKB rings, with a small aperture at the interaction region, doubled stored beam currents, and short expected lifetimes. Low-emittance, high-current electrons 
are delivered by employing a photo-cathode RF gun. High-current primary electrons for positron production are generated by a thermionic gun, and then high-current positrons 
are produced using a flux concentrator (FC) and large-aperture accelerating structures (LASs), which are then damped to low emittance through a DR.  
Design parameters of the injection beams are listed in Table~\ref{param2}. 

\subsection{Electron sources}

\begin{figure}[hbt]
\centering
\includegraphics*[width=80mm]{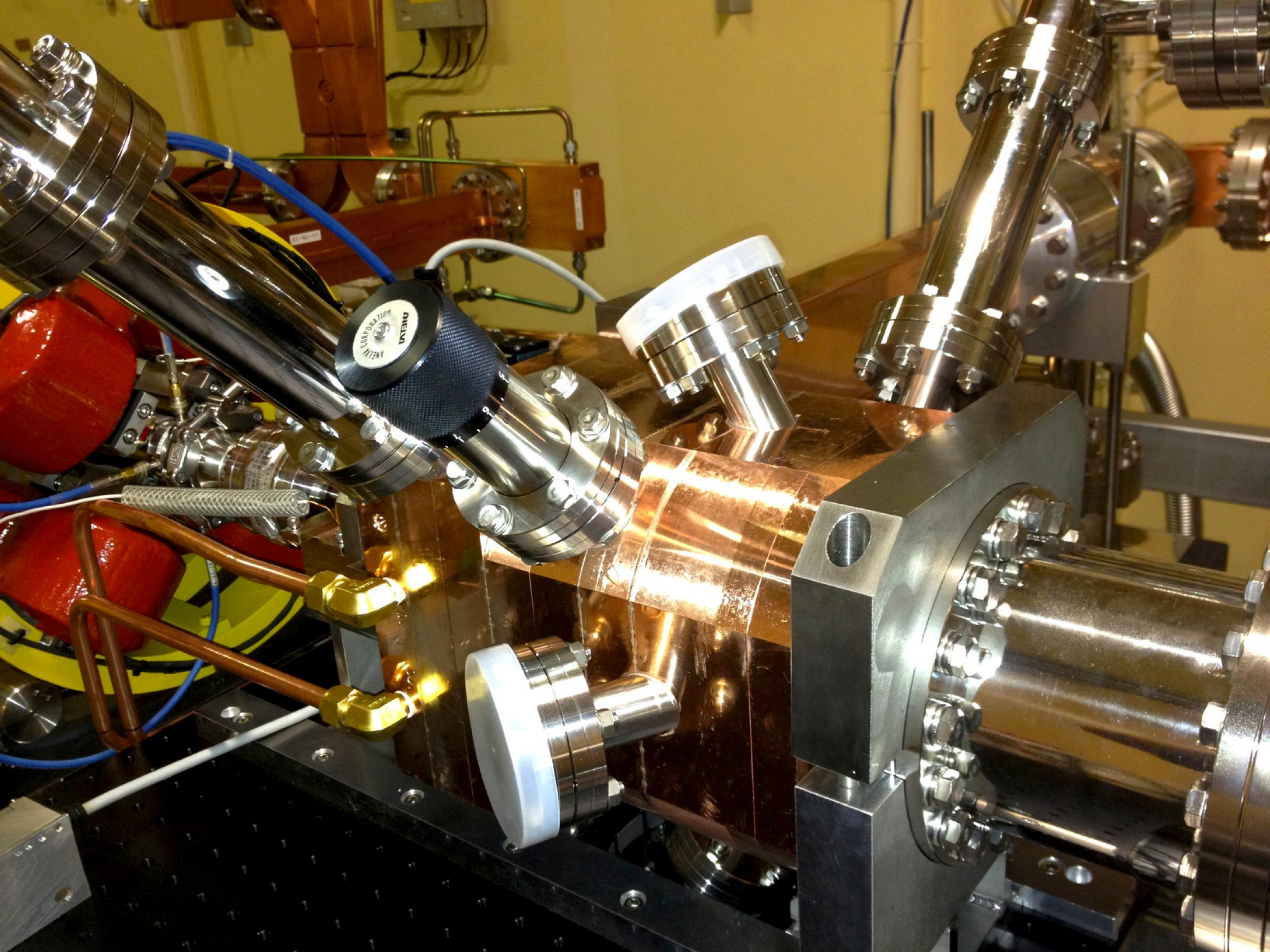}
\caption{Quasi-traveling-wave side-couple cavity with a high acceleration gradient.}
\label{qtwsc}
\end{figure}

A low-emittance electron beam source and its transport are indispensable to realizing SuperKEKB's nano-beam scheme for higher collision rates. Although DR is employed to reduce positron emittance, cost and space restrictions make the same solution infeasible for electrons. Thus, we have developed a photo-cathode high-current RF gun. Generating a high-charge electron bunch (up to 4 nC) is 
a challenging target. 
The primary target of the gun is a bunch charge of 4 nC and an emittance of 10 mm$\cdot$mrad to allow for minor emittance blow-up along the linac. Each component of the RF gun, such as the laser, photo cathode, and cavity, was examined carefully for stable long-term operation. A laser system with a Yb-doped fiber oscillator, a fiber amplifier, a thin-disk multipass amplifier, and two-stage frequency doublers was installed to 
realize high-power, shaped laser pulses~\cite{ipac17-laser}. 
A combination was chosen for a baseline, with an Nd:YAG laser for higher power, ${\rm {Ir_5}Ce}$ cathode for longer lifetime and reasonable quantum efficiency~\cite{ipac14-irce}, and a quasi-traveling-wave side-couple cavity (QTWSC) 
for higher accelerating gradient and focusing (Fig.~\ref{qtwsc})~\cite{eefact16-qtwsc}. A beam up to 4.4 nC was successfully transferred to the end of the linac with this combination.  

However, to deliver an electron beam with the required characteristics, space-charge-effect mitigation by a longer bunch length (30 ps) and energy-spread mitigation by a rectangular bunch shape should be manipulated carefully (temporal manipulation). Another laser station with a Yb:YAG thin-disk regenerative amplifier or multipass amplifier will be developed to achieve such a high-power, shaped laser pulses~\cite{linac14-gun}. 
A cut-disk structure cavity was also examined for front laser injection. The initial beam commissioning has been performed with a lower beam charge. 

\subsection{Positron generator enhancement}

\begin{figure}[hbt]
\centering
\includegraphics*[width=100mm]{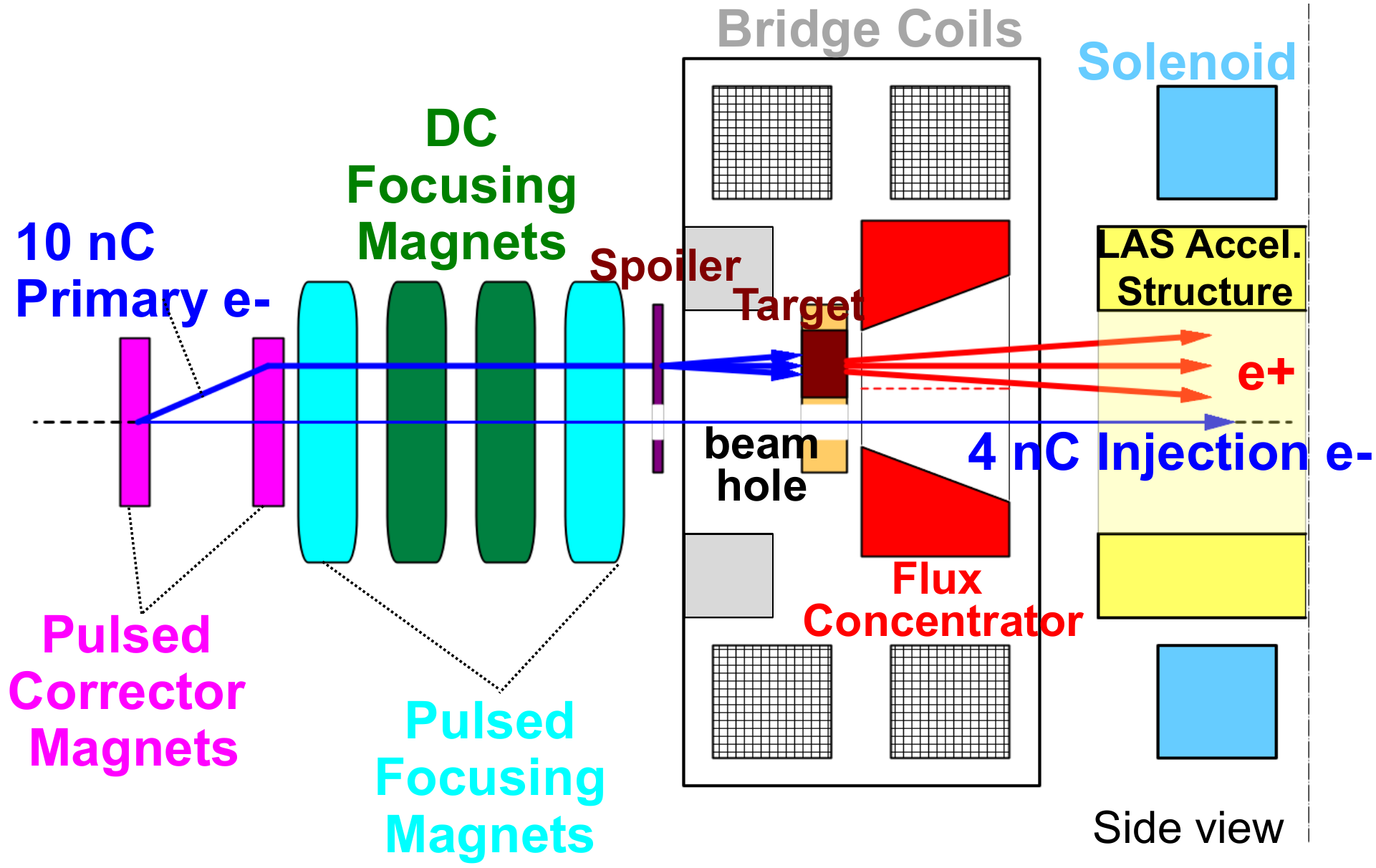}
\caption{Positron target with a hole beside is attached in front of the flux concentrator followed by LAS accelerating structures and solenoid coils.}
\label{target}
\end{figure}

The positron beam generated in the KEKB injector was approximately ${\rm 0.8 nC \times 2}$ bunches at 50 Hz. It should be enhanced 
to a dual-bunch 4 nC beam in stages, because SuperKEKB has the doubled stored beam current and a much shorter beam lifetime of several minutes. 
A high-charge positron bunch 
is generated with a conventional 14-mm thick tungsten target, and 
is captured by employing an FC and LASs with velocity bunching, followed by a series of solenoid and focusing magnets; this is shown in Fig.~\ref{target}~\cite{ipac14-pos}. As the generated beam emittance is large, it 
is damped by employing a DR at 1.1 GeV.  

One of the keys to increasing the positron yield is the development of a focusing pulsed coil just after the positron generation target. While an air-core pulsed coil was employed in KEKB, it was replaced by a pulsed FC with a high field of 3.5 T in SuperKEKB. The FC is a pulsed solenoid composed of a primary coil and a copper cylinder with a conical hole inside. Induced eddy current flows through a thin slit to an inner surface and generates a strong field. The achievable field strength is mainly determined by the hole diameter and primary pulsed current. 
The FC was fabricated at KEK 
and was operated at a peak current of 6 kA --- half of the nominal current --- without any issue. 
A 14-mm thick positron target 
is attached in front of the FC, followed by LAS accelerating structures, solenoid coils and a hundred of focusing magnets. A hole beside the target is used to pass the electron beam to be injected into the HER. 

A thermionic gun is utilized to deliver a high-charge 10 nC primary electron beam for positron generation. The beam optics were analyzed and corrected at the 24$^{\circ}$ merger beamline with RF guns.

Simulation studies from the thermionic electron gun to the DR, including the FC and LASs, were performed to optimize beam optics parameters with the target position, magnetic field, electric field, and collimator positions that lead to higher positron capturing ratios and lower beam loss.  
A target protection scheme, including a beam spoiler and a beam loss monitor, was also developed to prevent the energy deposit from exceeding a known limit~\cite{ipac14-fc}.

\subsection{Emittance preservation}

If a beam bunch accelerated in the structure is offset from the center, the generated wakefield induces a transverse force to the tail, and the projected emittance can be very large. According to a simulation, including the random alignment errors of the accelerating structures and the quadrupole magnets, the beam may easily have a 100-times larger emittance. 

\begin{figure}[hbt]
\centering
\includegraphics*[width=80mm]{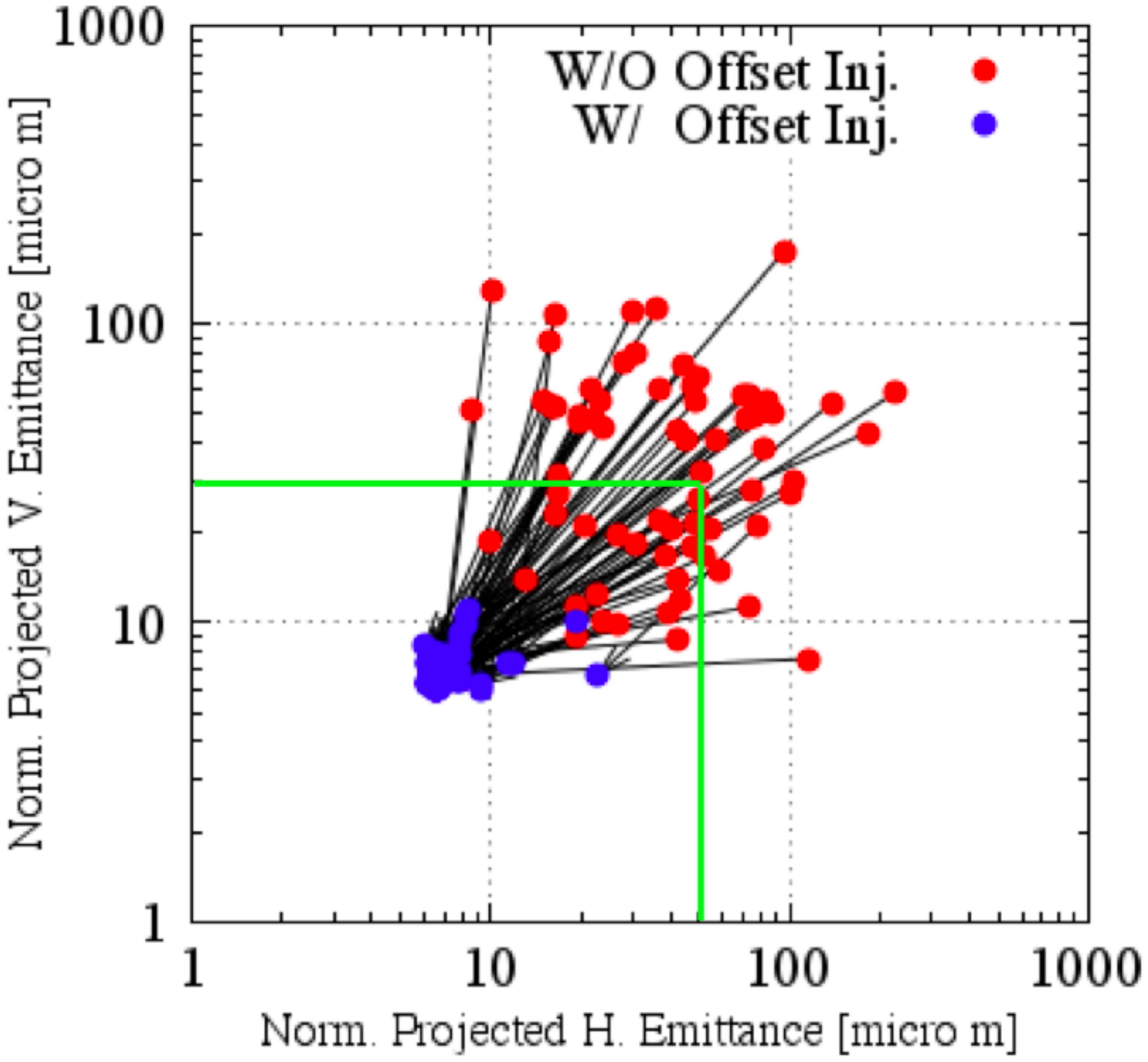}
\caption{Blow-up emittances without beam offset adjustment (red), and controlled recoveries of emittances (blue) within tentative target emittance (green) applying adjusted initial beam positions and angles for 100 random simulation orbits.}
\label{offset}
\end{figure}

Suppressing this effect requires mechanical alignment of the quadrupole magnets and accelerating structures.  
The simulation study suggests it is possible to deliver a required beam only if their alignment is less than 300 $\mu$m (rms) for an overall 600-m injector and 100 $\mu$m (rms) for a short segment, as shown in Fig.~\ref{offset}.
The beam orbit should be stabilized by finding a low-emittance condition empirically~\cite{ipac16-emitt}. 
In the method, the choice of an orbit can cancel the longitudinal bunch deformation. To that end, the angular accuracy of the beam steering coil must be less than 1 ${\mu}$rad~\cite{ipac15-emit}. A beam position read-out system of precision less than 10 $\mu$m is also developed to support the method~\cite{ibic15-bpm}.
 
An alignment of devices down to 100 ${\mu}$m locally and 300 ${\mu}$m globally is desirable along the 120-m and 480-m straight beamlines. Local measurements were taken using a conventional laser tracker. However, as the linac tunnel is narrow, global measurement was challenging. Thus, a new long-baseline laser beam was developed, which was stabilized by a precise angle adjuster, guided through vacuum pipes under the accelerator girders, and measured by position-sensitive photo detectors. 
The measurement precision was 60 ${\mu}$m over a 480-m straight beamline~\cite{iwaa14-align}.  
The girders for the quadrupole magnets were newly developed and installed with a precise alignment mechanism down to a micron.


\subsection{Simultaneous injection}

The newly installed equipment and monitors were designed to operate at 50 Hz. They 
are operated via event-based, global, and synchronized controls to inject beams with different properties into four separate storage rings simultaneously~\cite{ipac10-feedback,ipac14-event}.  

\begin{figure}[hbt]
\centering
\includegraphics*[width=80mm]{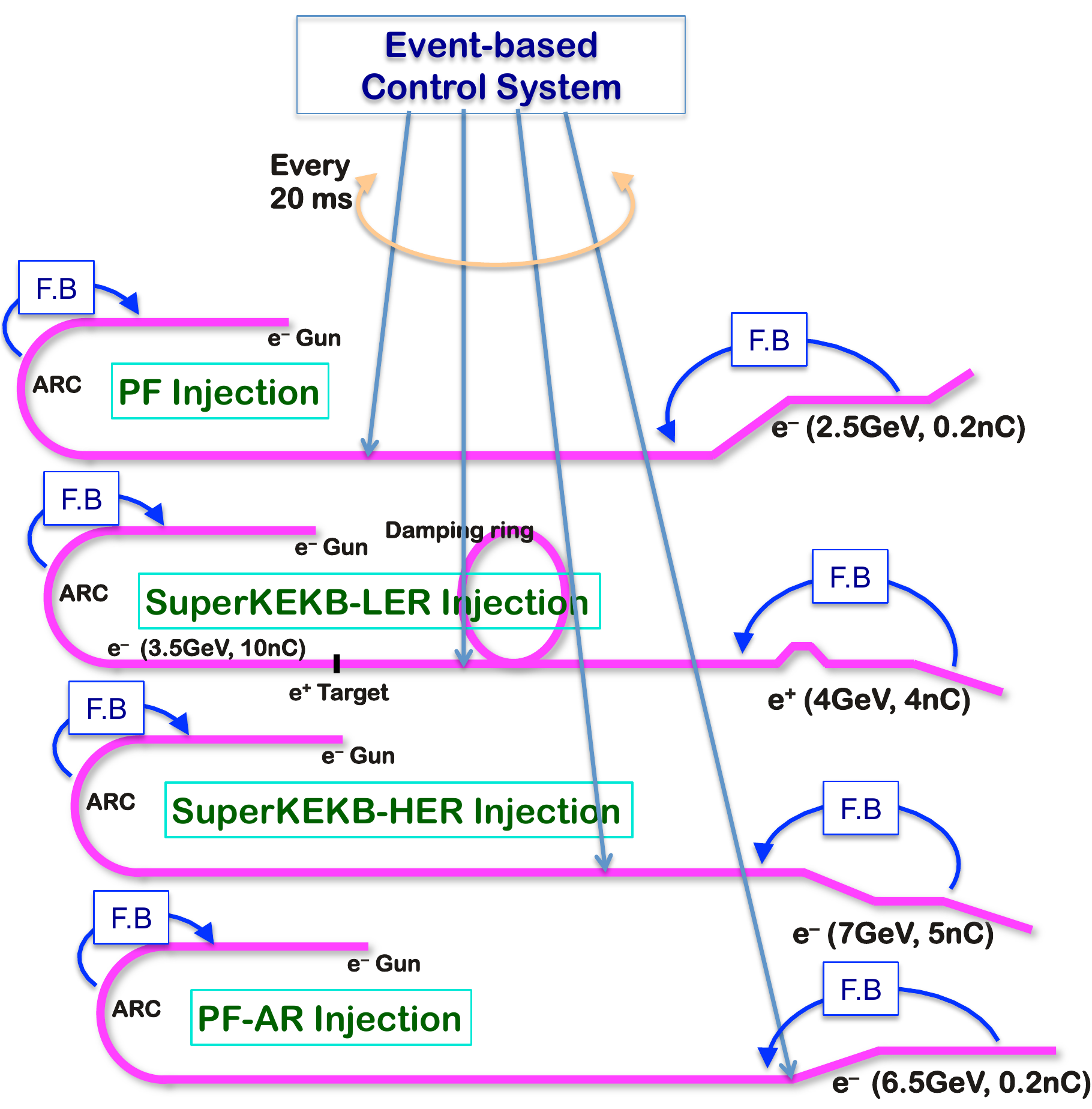}
\caption{Single linac behaves as four virtual accelerators to inject their beams into four separate storage rings.}
\label{va}
\end{figure}

A single injector linac would behave as four independent virtual accelerators (VAs) with hundreds of independent parameters modulated pulse-by-pulse at 50 Hz (Fig.~\ref{va}).  
A new beam transport line for PF-AR direct injection prevents interference between the PF-AR and SuperKEKB HER, which shared the same beam transport line in the KEKB project~\cite{ipac17-pfar}. 

In KEKB, pulse-to-pulse modulation (PPM) of injection beams was performed with moderate beam optics, and fine optics matching was performed at corresponding beam transport lines~\cite{ipac10-feedback}. However, for SuperKEKB injection, beam orbit and optics management is necessary within a precision of 100~${\mu}$m to suppress emittance blow-up~\cite{ipac16-emitt}.  


Compact power supplies were developed for quadrupole magnets with advanced design specifications of 1 mH, 330 A, and 340 V, and a 2 ms pulse width with up to 75\% energy recovery from the magnetic coils. The same pulsed power supplies are used to drive bending magnets to switch beams from a thermionic gun and RF guns. Power supplies for steering magnets were also designed for 3 mH and 10 A. 

Those power supplies were designed for pulses with a 0.5 ms flat top, and must be triggered 3 ms before the beam. Such timing signals could be directly generated for SuperKEKB beams because the linac and rings are tightly synchronized. Generating timing signals with a long delay for light sources, however, is complicated, because those rings compensate for the circumferences by modulating the RF frequencies, and they are injected based on timing coincidence and feedforward algorithm. 

Pulsed magnets of 30 quadrupoles, two bendings, and 36 steerings were mass-produced and installed during a five-month shutdown in 2017.  
The pulsed magnets were examined 
and confirmed to satisfy the specification of 0.1\% (rms) stability. 

\subsection{RF systems}\label{rfsystem}

Each of 60 high-power accelerating units is typically equipped with a pulse modulator, a S-band 50-MW klystron, a SLED-type energy doubler, and four 2-m long quasi-constant gradient accelerating structures.  
Those high-power RF modulators can operate at 50 Hz.  All of them shorten the charging time by 1 ms, in order for the injection bucket selection system to synchronize the linac and rings.  One-third of those modulators were replaced with compact inverter-type modulators to make space for new devices.  

A 60-kW driver klystron was originally employed to drive a group of 8 high power accelerating units.  However, many of units are required to operate independently in order to realize simultaneous injections. Thus, a 600-W solid state RF driver was developed with an FPGA-based digital IQ modulator.  They are directly connected with a master oscillator via temperature-stabilized optical links.  The master oscillator provides several RF frequencies for linac and rings with integer relations between them.  

As the beam quality is tightly dependent on the stability of the RF system, new RF monitors were developed.  The monitor is equipped with five sets of IQ detectors and fast ADCs, an FPGA, and direct optical links to the event and EPICS control systems.  About 70 monitors were installed to ensure the adequate RF stability for the beam specifications~\cite{pasj14-llrf}.  



\subsection{Beam instrumentation and controls}

Four storage rings should be filled simultaneously in top-up injection mode, as previously described.  
To this end, the linac should be operated with precise beam controls.  Dual-layer controls with EPICS and event-based systems have been enhanced to support beam operation with precise PPM at 50 Hz~\cite{ptep-cont}.  A VA is 
introduced to enable a single injector linac to behave as four independent VAs switched by PPM, in which each VA corresponds to one of four top-up injections into SuperKEKB's HER, LER, PF, and PF-AR. Each VA should be accompanied by independent beam feedback stabilization loops to preserve the orbit for low-emittance condition~\cite{ipac10-feedback}. 



A single-shot sliced emittance measurement would be possible, and we hope to operate on an unused bunch (a stealth bunch measurement) using VA and PPM controls. 



\section{Phase 1 commissioning}

\begin{figure}[hhh]
\centering
\includegraphics*[width=120mm]{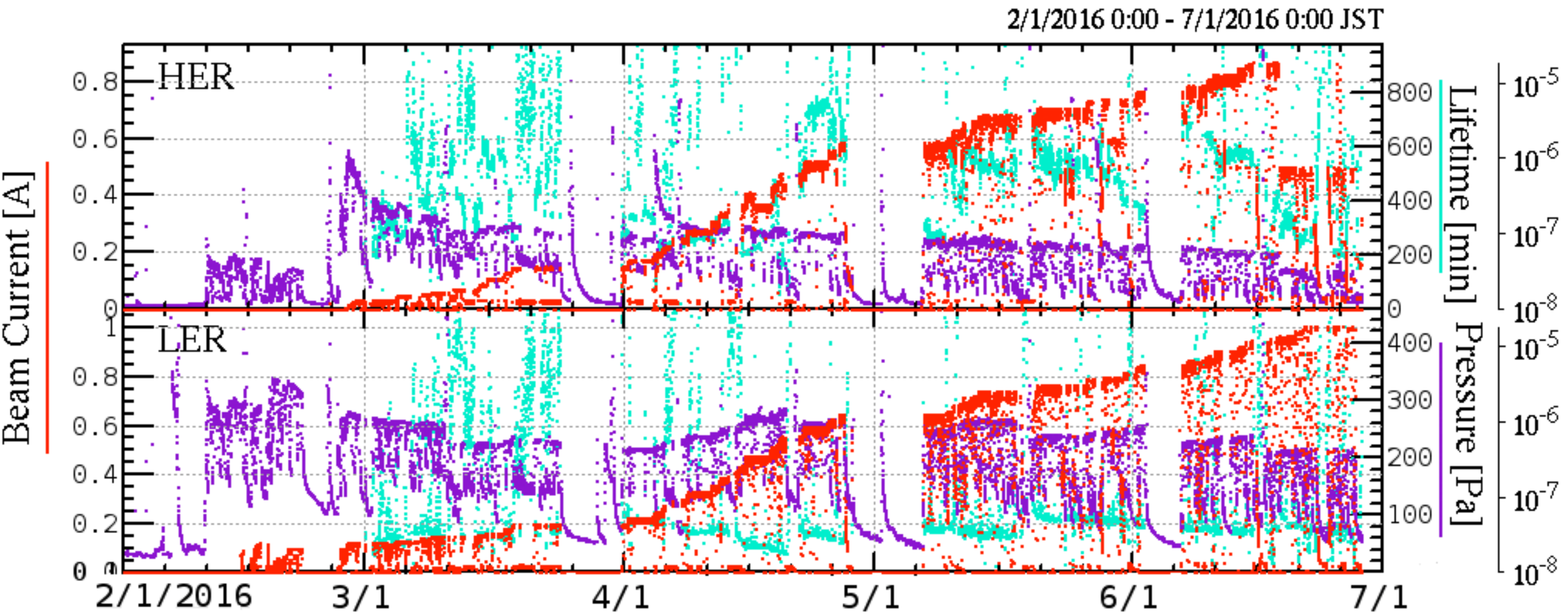}
\caption{History of the Phase 1 commissioning of SuperKEKB~\cite{MR_phase1_eefact2016}. Red points denote the beam current, purple the vacuum pressure, and cyan the beam lifetime.
}
\label{phase1}
\end{figure}

Phase 1 commissioning progressed smoothly, as shown in Fig.\ref{phase1}~\cite{MR_phase1_eefact2016}. 
Hardware systems (including newly introduced components) essentially worked as designed.
With the steady progress of the vacuum scrubbing, 
the beam currents reached 1.01 A and 0.87 A, average pressures of $\sim$1$\times10^{-6}$ Pa and $\sim$2$\times10^{-7}$ Pa, and beam doses of 780 A$\cdot$h and 660 A$\cdot$h  in the LER and HER, respectively~\cite{MR_vac_jvstA35}.
The beam doses fully satisfied the conditions for installing Belle II, which required reaching greater than 360 A$\cdot$h  (0.5 A $\times$ one month).

The new beam pipes with an antechamber structure and TiN coating worked well to suppress the ECE in the LER. 
However, we found that the aluminum bellows chambers, where no TiN coating was applied, significantly contributed to the increase in electron cloud density, which caused the ECE.
To cure this problem, permanent magnets with yokes were attached to the bellows chambers to form an axial magnetic field. Then the ECE was effectively suppressed~\cite{MR_vac_prst19_VacCommis}. 
In drift space, even with antechambers and TiN coating, electron clouds still formed without longitudinal magnetic fields;
hence, before Phase 2 commissioning, permanent magnets with yokes were installed in the drift space of all arc sections and the Tsukuba straight section for higher beam currents. 

\section{Summary}

Steady progress is being made in upgrading KEKB to SuperKEKB, a collider that will further extend the luminosity frontier for elementary particle physics. After 5.5 years of large-scale construction to upgrade the LER, HER, and linac, Phase 1 beam commissioning of SuperKEKB was successfully conducted in 2016. Subsequently, the QCS and Belle II detector have been installed, and renovation of the IR and construction of the new positron DR 
were completed for the start of Phase 2 beam commissioning, scheduled for early 2018.

\section*{Acknowledgement}

The authors would like to thank contributions of the physicists, engineers and students involved in cooperative research and development of the accelerator components incorporated in SuperKEKB. The authors also would like to thank researchers for participating in commissioning of SuperKEKB and fruitful discussions. 




\end{document}